\documentclass{aa}
\usepackage{graphicx}
\usepackage{natbib}
\usepackage{amsmath}
\bibpunct{(}{)}{;}{a}{}{,}
\newlength{\newsp}
\newcommand{\Ab}{\boldsymbol{A}}
\newcommand{\Bb}{\boldsymbol{B}}
\newcommand{\ab}{\boldsymbol{a}}
\newcommand{\bb}{\boldsymbol{b}}
\newcommand{\fb}{\boldsymbol{f}}
\newcommand{\abt}{\boldsymbol{\tilde a}}
\newcommand{\bbt}{\boldsymbol{\tilde b}}
\newcommand{\Pbt}{\boldsymbol{\tilde P}}

\newcommand{\Fb}{\boldsymbol{F}}
\newcommand{\gb}{\boldsymbol{g}}
\newcommand{\gbh}{\boldsymbol{\widehat{g}}}

\newcommand{\gbt}{\boldsymbol{\tilde g}}
\newcommand{\Gb}{\boldsymbol{G}}
\newcommand{\Hb}{\boldsymbol{H}}

\newcommand{\Lb}{\boldsymbol{L}}
\newcommand{\vb}{\boldsymbol{v}}

\newcommand{\Rb}{\boldsymbol{R}}

\newcommand{\ub}{\boldsymbol{u}}

\newcommand{\Pb}{\boldsymbol{P}}

\newcommand{\wb}{\boldsymbol{w}}

\newcommand{\zb}{\boldsymbol{z}}

\newcommand{\fmatb}{\boldsymbol{{\cal F}}}

\newcommand{\rddots}{\tiny

\begin{array}{ccc} & & \cdot \\ & \cdot & \\ \cdot & & \end{array}
}

\begin{document}
\title{Digital Deblurring of CMB Maps II: \\ Asymmetric Point Spread Function}

   \author{R. Vio\inst{1}
          \and
            J.G. Nagy\inst{2}
          \and
          L. Tenorio\inst{3}
          \and
            P. Andreani\inst{4}
          \and
            C. Baccigalupi\inst{5}
          \and
          W. Wamsteker\inst{6}
          }

   \offprints{R. Vio}

   \institute{Chip Computers Consulting s.r.l., Viale Don L.~Sturzo 82,
              S.Liberale di Marcon, 30020 Venice, Italy\\
              ESA-VILSPA, Apartado 50727, 28080 Madrid, Spain\\
              \email{robertovio@tin.it}
         \and
                  Department of Mathematics and Computer Science, Emory University,
                        Atlanta, GA 30322, USA. \\
              \email{nagy@mathcs.emory.edu}
         \and
              Department of Mathematical and Computer Sciences, Colorado School
                  of Mines, Golden CO 80401, USA \\
              \email{ltenorio@Mines.EDU}
         \and
              Osservatorio Astronomico di Padova, vicolo dell'Osservatorio 5,
                  35122 Padua, Italy \\
              \email{andreani@pd.astro.it}
         \and
                   SISSA/ISAS, Via Beirut 4, 34014 Trieste, Italy \\
              \email{bacci@sissa.it}
         \and
             ESA-VILSPA, Apartado 50727, 28080 Madrid, Spain\\
             \email{willem.wamsteker@esa.int}
             }

\date{Received .............; accepted ................}

\abstract{In this second paper in a series dedicated to developing efficient
numerical techniques for the deblurring Cosmic Microwave Background (CMB)
maps, we consider the case of asymmetric point spread functions (PSF). Although
conceptually this problem is not different from the symmetric case, there are important
differences from the computational point of view because it is no longer possible to use some
of the efficient
numerical techniques that work with symmetric PSFs. We present procedures that permit the use
of efficient techniques even when this condition is not met. In particular, two methods are considered:
a procedure based on a Kronecker approximation technique that can be implemented with the
numerical methods used with symmetric PSFs but that has the limitation of requiring only
mildly asymmetric PSFs. The second is a variant of the classic Tikhonov technique that works even with
very asymmetric PSFs but that requires discarding the edges of the maps. We
provide details for efficient implementations of the algorithms. Their performance is tested on simulated
CMB maps.
}

\titlerunning{Digital Deblurring of CMB maps}
\authorrunning{R. Vio, J.G. Nagy, L. Tenorio et al.}
\maketitle

\section{Introduction}

In the first paper in the series \citet{vio03} (VNT) have stressed the advantage of
deblurring small patches
of the sky in cosmic microwave background (CMB) studies: first, it helps to
recover high frequencies smoothed
out by the instrument's PSF. Second, a better understanding of sky
emissions, from foregrounds in particular,
is achieved if multifrequency sky maps are compared on a common resolution.
Third, some map-based component
separation algorithms, such as independent component analysis \citep{bac00, mai02},
require input maps with similar level
of degradation. Furthermore, although the aim of satellite missions such as
{\it Planck} and {\it MAP} is to obtain full sky maps of the CMB, the strength of the CMB
over other backgrounds or contaminating
sources will vary over the sky. As a result, even a successful separation
of the components contributing to the microwave radiation will provide results of inhomogeneos
quality. Therefore, even if some characteristics of
CMB are estimated on full sky maps,
it will be important to check these results on smaller sky patches where
CMB largely dominates the other components (i.e., no component separation is necessary as,
e.g., at high Galactic latitude and at high observing frequency) and data
are free from instrumental and/or observational problems. We stress that a gain in resolution is 
always possible also when maps of different resolutions are composed together, for example, through the
method described in \citet{teg99}.

VNT suggested a deblurring approach based on Tikhonov regularization whose computational
cost is comparable to that of classic frequency domain methods but that leads to more reliable and
stable deblurring estimates. These new efficient implementations make the Tikhonov technique
a promising tool for
deblurring CMB maps. The main limitation of the method is the requirement of symmetric PSFs, which prevents
its application in the general experimental context with no restriction on the form of the PSF.
This point is especially important in CMB experiments where an asymmetric PSF can alter
the results of the analyses, for example, by distorting the estimated angular power spectrum or
by altering the measure of the degree of nonGaussianity of the maps.
Since some experimental situations are not well controlled, one should develop algorithms that
can cope with the {\it worst possible scenario}; it is thus important to develop deblurring methods for
asymmetric PSFs.

In Sect.~\ref{sec:formal} we formalize the problem and propose two solutions in Sects.~\ref{sec:kronecker}
and \ref{sec:window}: the first one is a very efficient method but requires only mildly asymmetric
PSFs, whereas the second, although less efficient, is not limited by the particular form of the observing beam.
A modification of the second method is considered in Sect.~\ref{sec:extension}. The results of numerical simulations
to test the performance of the different methods are presented in Sect.~\ref{sec:numerical}.
In Sect.~\ref{sec:final} we close with final comments and conclusions.

\section{Formalization of the problem} \label{sec:formal}

We make use of the same formalism adopted in
VNT.
When a two-dimensional object $f(\xi,\eta)$ is
observed through an optical invariant (linear) system, it is seen as an image $g(x,y)$,
\begin{equation} \label{eq:model}
g(x,y) = \int\limits_{-\infty}^{+\infty} \int\limits_{-\infty}^{+\infty} h(x-
\xi,y-\eta) f(\xi,\eta)~d\xi~d\eta,
\end{equation}
where the space-invariant {\it point-spread function} (PSF) $h(x-\xi,y-\eta)$
represents the blurring action of the optical instrument. This model is only theoretical,
in practical applications
we only have discrete noisy observations of the image $g(x,y)$, which
we model as a discrete linear system
\begin{equation} \label{eq:modeld}
\gb = \Hb \fb + \zb,
\end{equation}
where: $\gb = {\rm vec} (\Gb)$ and $\fb = {\rm vec} (\Fb)$ are one-dimensional
column arrays containing, respectively, the observed image $\Gb$ and the true images
$\Fb$ in {\it stacked} order, $\zb$ is an array containing the noise contribution
(assumed to be additive), and $\Hb$ is a matrix that represents the discretized blurring operator.

There are two problems in obtaining an estimate of $\fb$ from
$\gb$: the size of the matrix $\Hb$, which is large even for
moderate size images, and the very ill-posed nature of the problem. VNT proposed a
deblurring method for CMB applications that is efficient and numerically stable; it is a Tikhonov
regularization
approach where the estimate $\fb_{\lambda}$ of $\fb$ is defined as
\begin{equation} \label{eq:tikhonov}
\fb_{\lambda} = {\rm argmin}\left(\, \Vert\, \Hb \fb - \gb \,\Vert_2^2 + \lambda^2 \Vert\,
\Lb \fb \,\Vert_2^2\, \right),
\end{equation}
with $\lambda$ a scalar regularization parameter. $\Lb$ is
often the identity matrix or a discrete derivative operator of some order.

An additional problem is the selection of boundary conditions (BC) to account for
data outside the field of view. VNT found that better deblurring estimates of CMB maps were
obtained with reflexive BC. This choice leads to reliable and stable regularization parameter estimates,
and helps suppress spurious features such as Gibbs oscillations. Periodic and zero BC
impose edge discontinuities which bias the image Fourier coefficients and affect the
reliability of the regularization parameter estimates obtained through generalized cross-validation
(GCV).

On the other hand, efficient implementations of Tikhonov deblurring with reflexive BC
require symmetric PSF (although not necessarily separable), i.e., $h(x,y) = h(-x,y) = h(x,-y)= h(-x,-y)$.
Since the PSF may be asymmetric in some practical applications (there are indications that this may be the case
for {\it PLANCK}'s optics), we consider efficient implementations of Tikhonov
deblurring that can be used in this case. In particular, in Sect.~\ref{sec:kronecker} we consider a very efficient
method based on reflexive BC that, however, works only when the PSF is slightly asymmetric. In
Sect.~\ref{sec:window} a less efficient method, based on periodic BC, is presented; its performance
is not sensitive to the specific form of the PSF. For this last case we have to modify the traditional
GCV to provide regularization parameter estimates in the presence of edge discontinuities.

\section{Kronecker approximation} \label{sec:kronecker}

One possible alternative when the PSF is not symmetric is to
determine if the PSF is at least separable; that is,
\begin{equation}\label{eq:h}
  h(x,y) = h_x(x)h_y(y)\,.
\end{equation}
If this is the case, then the corresponding $n^2 \times n^2$ matrix $\Hb$ can be written as
\begin{equation}
  \Hb = \Ab \otimes \Bb\,,
\end{equation}
where $\Ab$ and $\Bb$ are $n \times n$ matrices, and $\otimes$ is used
to denote a Kronecker product
$$
  \Ab \otimes \Bb =
  \left(
    \begin{array}{cccc}
      a_{11} \Bb & a_{12} \Bb & \cdots & a_{1n} \Bb \\
      a_{21} \Bb & a_{22} \Bb & \cdots & a_{2n} \Bb \\
      \vdots & \vdots & & \vdots \\
      a_{n1} \Bb & a_{n2} \Bb & \cdots & a_{nn} \Bb
    \end{array}
  \right)\,.
$$
For such a structured matrix, algorithms can be implemented
efficiently. The cost is $O(n^3)$, which is slightly more than the
$O(n^2\log n)$ required of transform-based methods, but is still very
reasonable for large images; see VNT for further details.
Therefore, we should exploit this structure if we can recognize that the blur is separable.

The key point that permits the development of an efficient algorithm is
that if $\Pb$ is an $n \times n$ array of pixels representing the
observed PSF and the blur is separable, then $\Pb$ is a rank-one
matrix; that is, $\Pb$ can be written as an outer product
\begin{equation}
  \Pb = \ab \bb^T,
\end{equation}
where $\ab$ and $\bb$ are $n\times 1$ vectors. For
reflexive boundary conditions the matrices $\Ab$ and $\Bb$
(where $\Pb$ is such that $p_{ij}$ is the center of the PSF)
are Toeplitz-plus-Hankel of the form
\begin{equation}
  A = \left( \begin{array}{ccccc}
                 a_i & \cdots & a_1 & & \\
               \vdots & \ddots & & \ddots & \\
                 a_n & & \ddots & & a_1 \\
                      & \ddots & & \ddots & \vdots \\
                      & & a_n & \cdots & a_i
             \end{array}
       \right) +
       \left( \begin{array}{ccccc}
                 a_{i+1} & \cdots & a_n & \\
                 \vdots & \rddots & & \\
                   a_n & & & a_1 \\
                         & & \rddots & \vdots \\
                         & a_1 & \cdots & a_{i-1}
             \end{array}
       \right)\,,
\end{equation}
and
\begin{equation}
  B = \left( \begin{array}{ccccc}
                 b_j & \cdots & b_1 & & \\
               \vdots & \ddots & & \ddots & \\
                 b_n & & \ddots & & b_1 \\
                      & \ddots & & \ddots & \vdots \\
                      & & b_n & \cdots & b_j
             \end{array}
       \right) +
       \left( \begin{array}{ccccc}
                 b_{j+1} & \cdots & b_n & \\
                 \vdots & \rddots & & \\
                   b_n & & & b_1 \\
                         & & \rddots & \vdots \\
                         & b_1 & \cdots & b_{j-1}
             \end{array}
       \right)\,.
\end{equation}
If the PSF if effectively separable,
then $\ab$ and $\bb$ can be obtained, respectively, by discretizing
the functions $h_x(x)$ and $h_y(y)$ in Eq.~(\ref{eq:h}).

Of course, these arguments do not hold when the PSF is not separable
since it cannot be expressed exactly as an outer product
of two vectors. However, if the PSF is only approximately non-separable,
then it is possible to work with a separable approximation of $\Pb$.
The most natural approach is to make use of the singular value decomposition
(SVD) of $\Pb$:
\begin{equation}
   \Pb = \sum_{i=1}^r \sigma_i \ub_i \vb_i^T,
\end{equation}
where the singular values $\sigma_i$ satisfy
$\sigma_1 \geq \cdots \geq \sigma_r > 0$,
$r$ is the rank of $\Pb$, and the singular
vectors $\ub_i$ and $\vb_i$ satisfy
\begin{equation}
  \ub_i^T\ub_j = \vb_i^T\vb_j =
  \left\{ \begin{array}{ll}
            1 & \mbox{ if } i = j \\
            0 & \mbox{ if } i \neq j\,.
          \end{array}
  \right.
\end{equation}
The best rank-one approximation of $\Pb$ is given by \citep[e.g., ][]{GoVa96}
\begin{equation}
  \Pb \approx \sigma_1 \ub_1 \vb_1^T\,.
\end{equation}
Thus, we may take $\ab = \sqrt{\sigma_1}\ub_1$ and
$\bb = \sqrt{\sigma_1}\vb_1$, which implies
\begin{equation}
\label{eq:EasyApprox}
   \Pb \approx {\bf a}{\bf b}^T \quad \mbox{ and } \quad
   \Hb \approx \Ab \otimes \Bb\,.
\end{equation}
Although this often works well in practice, it is generally
preferable, and mathematically more satisfying,
to have an {\em optimal} approximation. Specifically,
we would like to find a Kronecker product $\Ab \otimes \Bb$
that minimizes
\begin{equation}
  \min||\Hb - \Ab \otimes \Bb||\,,
\end{equation}
where $||\cdot||$ is a chosen norm,
and the minimization is done over all Kronecker products $\Ab \otimes \Bb$.
Approximations in the case of the Frobenius norm ( $\|\cdot\|_F$)
have recently received a lot of attention. \cite{VLoPi93}
developed the idea for general matrix approximations, which was made
computationally efficient for image restoration problems with
zero boundary conditions by \cite{KaNa00}.
For reflexive boundary conditions, the optimal approximation
can be computed using the following theorem, which was recently
established by \cite{NaNgPe02}.
\begin{theorem}
Let $\Pb$ be an $n \times n$ PSF. For reflexive boundary conditions:
$$
  ||\Hb - \Ab \otimes \Bb||_F = ||\tilde{\Pb} - \abt \bbt^T||_F
$$
where $\Pbt = \Rb\Pb\Rb^T$, $\abt = \Rb\ab$, $\bbt = \Rb\bb$ and
$\Rb$ is the Cholesky factor of the $n \times n$ symmetric Toeplitz
matrix with first row
$
  [ \begin{array}{ccccccccc}
           n & 1 & 0 & 1 & 0 & 1 & 0 & 1 & \cdots
         \end{array}
  ]$ .
\end{theorem}

\vspace{2mm}
\noindent
The proof of this theorem, which requires many tedious details,
is given in \cite{NaNgPe02}. Based on this theorem, an algorithm
for constructing the optimal Kronecker product approximation
is as follows:

\noindent
{\bf Algorithm:}
To construct the approximation $\Hb \approx \Ab \otimes \Bb$:
\begin{itemize}
\item Compute $\Rb$
\item Construct $\Pb_r = \Rb\Pb\Rb^T$
\item Compute the SVD: $\Pb_r = \sum \sigma_k \ub_k \vb_k^T$
\item Construct the vectors:
$$
  \ab = \sqrt{\sigma_1}~\Rb^{-1}\ub_1 \quad \mbox{and} \quad
  \bb = \sqrt{\sigma_1}~\Rb^{-1}\vb_1
$$
\item Construct the matrices $\Ab$ and $\Bb$ from
$\ab$ and $\bb$ (as described above).
\end{itemize}
If the PSF image array is of size $m \times m$, then the cost of
constructing this optimal Kronecker product approximation is
only $O(m^3)$, which is relatively cheap if
the width of the PSF is small compared
to the dimension of the blurred image (i.e., $m \ll n$).
We use this scheme in our computations
because it produces a provably optimal approximation, and
because it is
essentially equivalent in cost to the straight forward approach
given in Eq.~(\ref{eq:EasyApprox}).

\section{Image windowing} \label{sec:window}

The Kronecker approximation method performs poorly when the PSF is very asymmetric. We now consider a
deblurring method based on periodic BC which can be implemented using the fast Fourier transform.

\begin{figure}
        \resizebox{\hsize}{!}{\includegraphics{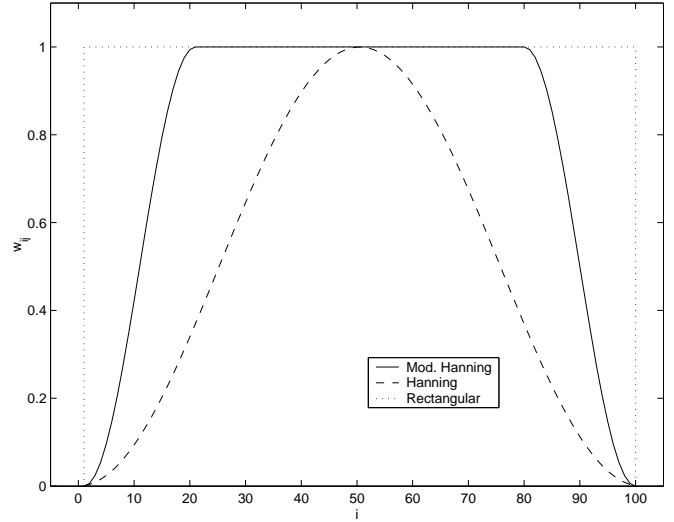}}
        \caption{Slice comparison of the two-dimensional classical Hanning and rectangular windows
        with the modified Hanning window ($N_w=20$).}
        \label{fig:windows}
\end{figure}

\begin{figure}
        \resizebox{\hsize}{!}{\includegraphics{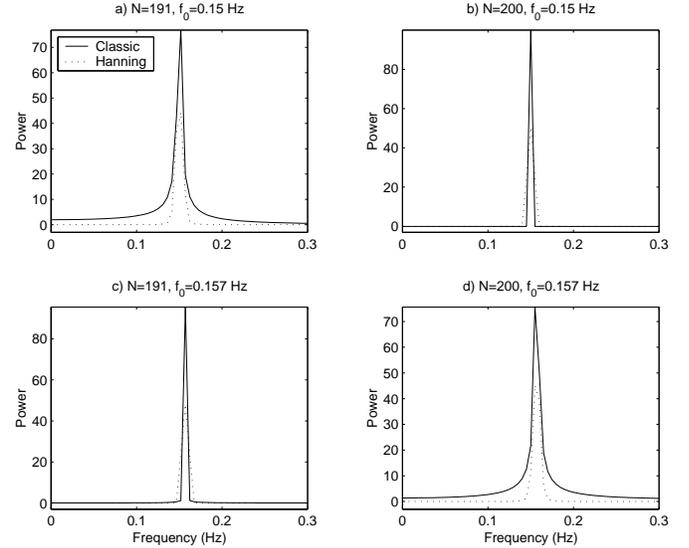}}
        \caption{Power-Spectra of four different realizations of the process $x[t] = \sin(2 \pi f_0 t)$,
        $t=0,1,\ldots, N$ for $f_0= 0.150$ and $0.157~{\rm Hz}$, $N=191$ and $200$. The true power-spectrum
        (not shown) is a $\delta$-function centered at $f_0$.}
        \label{fig:win}
\end{figure}

A problem with periodic BC is the effect of edge discontinuities in regularization parameter estimates. This effect
can be reduced by considering an average GCV defined by an estimated spectrum. More precisely, the GCV in the
spectral domain is (see VNT)
\begin{equation} \label{eq:gcv}
  {\rm GCV}(\lambda) =
                 n
                 \sum_{i=1}^{n}
                 \left(\frac{ \delta_i^2\,\|\hat{g}_i\|}
                       {\sigma_i^2+\lambda^{2}\delta_i^2}
                 \right)^{2}
                  /
                  \left(
                         \sum_{i=1}^{n}
                         \frac{\delta_i^2}
                              {\sigma_i^2+\lambda^2 \delta_i^2}
                  \right)^2
\end{equation}
where: $\{ \sigma_i \}$ and $\{ \delta_i \}$ are the eigenvalues of $\Hb$ and $\Lb$, respectively, $\gbh=\fmatb\gb$,
with $\fmatb$ the bidimensional Fourier matrix, and $n$ is the number of pixels in the image. To reduce
edge effects on $\gbh$ we use an estimate of $\|\,\gbh\,\|^2$ (the spectrum) based on bidimensional windowing
of the Fourier transform: $\gbh$ is estimated via
$\gbt_1=\fmatb (\wb_1 \odot \gb)$, where `` $\odot$ ''
denotes element-wise matrix multiplication and $\wb_1$ is a bidimensional function (window) tapered
smoothly to zero at the image edges. A typical choice for $\wb_1$ is the {\it Hanning} window
(see Fig.~\ref{fig:windows}), but many other alternatives are available. The need for windowing is
evident in Fig.~\ref{fig:win}, which shows a simple example based on the realization of a pure sinusoidal process.
It is clear that without windowing the (classic) estimated power-spectrum of the signal strongly depends on the
the specific sampling pattern. Note the spread of power at all the frequencies,
visible in panels ${\rm (a)}$ and ${\rm (d)}$, which has deleterious effects in
regularization parameter estimates. Windowing stabilizes the power-spectrum estimates.

Once the parameter $\lambda$ is estimated, a second windowing is necessary to also reduce edge effects in the
deblurring stage, i.e., the deblurring operation has to be carried out on a windowed map
$\gbt_2=\fmatb (\wb_2 \odot \gb)$. In this step, however, $\wb_2$ should distort the map as little as possible.
A possible solution is a window that does not alter the image within a central subimage. We use the following
modification of the classic Hanning window
\begin{equation} \label{eq:window}
(w_2)_{ij} = \left\{
\begin{array}{ll}
0.25 \times \alpha \times \beta & 1 \le i, j \le N_w; \\
w_{(N-i+1)(N-j+1)} & N - N_w < i, j \le N; \\
1 & {\rm otherwise};
\end{array}
\right.
\end{equation}
where $\alpha = [1-\cos(\pi~(i-1)/N_w)]$, and $\beta = [1-\cos(\pi~(j-1)/N_w)]$ so that
the pixels in the central subimage are not modified and the image has continuous first derivatives
at the edges (see Fig.~\ref{fig:windows}). This window approaches the classic
two-dimensional Hanning window as $N_w \rightarrow N/2$ and tends to the rectangular
windows as $N_w \rightarrow 0$. The parameter $N_w$ thus determines the filtering
characteristics, in particular the frequency pass-band, of the window. Its ``optimal'' value depends on
many factors such as the noise level, the form of the PSF and the specific realization of the
process.

\section{Reflexive image extension} \label{sec:extension}

Another method that allows, at least in principle, the use of reflexive BC with asymmetric
PSFs and which does not require discarding any data, consists of extending the original image
$X$ according to the scheme
\begin{displaymath}
\begin{array}{ccc}
X_{rc} & X_r & X_{rc} \\
X_c & X & X_c \\
X_{rc} & X_r & X_{rc}
\end{array}
\end{displaymath}
where $X_c$ is obtained by ``flipping'' the columns of $X$, $X_r$ is obtained
by ``flipping'' the rows of $X$, and $X_{rc}$ is obtained by ``flipping'' the rows and
columns of $X$. Since the resulting image is periodic, periodic BC
can be used without introducing discontinuities. However the computational cost is higher
as the new image is nine times larger. A more feasible approach is to make $X_{rc}$, $X_{c}$,
and $X_{r}$ only part of the image, say bands of thickness $N_{{\rm ext}}$ of the
same order as the width of the PSF. Since the discontinuities introduced
by the BC are expected mainly at the edges, this image extension keeps
edge effects on a part of the image that is later discarded. This method, however, has the disadvantage
of not providing full reflexive
BC and thus it has to be implemented using a windowing operation similar to that presented
in Sect.\ref{sec:window} with $N_w = N_{{\rm ext}}$.


\begin{table*}[t]
\begin{center}
\begin{tabular}{ccccccccccccc}
\hline
\hline
\multicolumn{1}{c}{} &
\multicolumn{1}{c}{{\rm Wiener}} & \multicolumn{1}{c}{} &
\multicolumn{1}{c}{{\rm Ref. Ext.}} & \multicolumn{1}{c}{} &
\multicolumn{1}{c}{{\rm Tikhonov}} & \multicolumn{1}{c}{} &
\multicolumn{1}{c}{{\rm Wind. Ref. Ext.}} & \multicolumn{1}{c}{} &
\multicolumn{1}{c}{{\rm Wind. Tikh.}} & \multicolumn{1}{c}{} &
\multicolumn{1}{c}{{\rm Kron. Approx.}} \\
\cline{2-2} \cline{4-4} \cline{6-6} \cline{8-8} \cline{10-10} \cline{12-12} \\
${\rm S/N}$ & ${\rm rrms\, (\%)}$ & & ${\rm rrms\, (\%)}^{\mathrm{a}} $ & & ${\rm rrms\, (\%)}$ &
& ${\rm rrms\, (\%)}^{\mathrm{a}} $ & &
${\rm rrms\, (\%)}$ & & ${\rm rrms\, (\%)}^{\mathrm{a}}$ & \\
\hline
\hline
2 & $50.71 \pm 0.12$ & & $51.09 \pm 0.20$ & & $50.18 \pm 0.21$ & & $50.99 \pm 0.18$ & &
$49.99 \pm 0.20$ & & $52.37 \pm 0.37$ \\
10 & $45.64 \pm 0.07$ & & $52.01 \pm 0.39$ & & $47.08 \pm 0.28$ & & $50.90 \pm 0.37$ & &
$45.50 \pm 0.13$ & & $*$ \\
100 & $41.29 \pm 0.04$ & & $*$ & & $*$ & & $*$ & &
$42.83 \pm 0.19$ & & $*$ \\
\hline
\hline
\end{tabular}
\caption{Summary of the results obtained for a simulated sky map contaminated with $100$ different
realization of a white noise process at the
{\it PLANK}-LFI frequencies and axial ratio of the elliptical PSF equal to $1:1.3$ (see text).
The central $300 \times 300$ pixels of the images are considered.
The original maps consisted of $364 \times 364$ pixels, i.e., a border of $32$ pixels has been removed from
each side of the images, corresponding to about four times the dispersion of the PSF along the major axis.
$N_w = 32$ for the modified Hanning window used for the windowed reflexive expansion and the windowed
Tikhonov methods and for the calculation of the regularization parameter in the reflexive expansion and
Tikhonov methods (see text). All of the methods have adopted a discrete Laplacian for $\Lb$ and (
except for the Kronecker approximation) periodic BCs.
The Kronecker approximation uses reflexive BCs.
The relative {\it root mean square} ({\rm rrms}) is defined
as the ratio of the residual root mean square ({\rm rms}) to the rms of the true signal. The asterisk means
unstable results.\label{tbl:sim1}}
\end{center}
\vspace*{\newsp}
\begin{list}{}{}
\item[$^{\mathrm{a}}$] Calculated on full $364 \times 364$ pixel images.
\end{list}
\end{table*}
\begin{table*}[t]
\begin{center}
\begin{tabular}{ccccccccccccc}
\hline
\hline
\multicolumn{1}{c}{} &
\multicolumn{1}{c}{{\rm Wiener}} & \multicolumn{1}{c}{} &
\multicolumn{1}{c}{{\rm Ref. Ext.}} & \multicolumn{1}{c}{} &
\multicolumn{1}{c}{{\rm Tikhonov}} & \multicolumn{1}{c}{} &
\multicolumn{1}{c}{{\rm Wind. Ref. Ext.}} & \multicolumn{1}{c}{} &
\multicolumn{1}{c}{{\rm Wind. Tikh.}} & \multicolumn{1}{c}{} &
\multicolumn{1}{c}{{\rm Kron. Approx.}} \\
\cline{2-2} \cline{4-4} \cline{6-6} \cline{8-8} \cline{10-10} \cline{12-12} \\
${\rm S/N}$ & ${\rm rrms\, (\%)}$ & & ${\rm rrms\, (\%)}^{\mathrm{a}} $ & & ${\rm rrms\, (\%)}$ &
& ${\rm rrms\, (\%)}^{\mathrm{a}} $ & &
${\rm rrms\, (\%)}$ & & ${\rm rrms\, (\%)}^{\mathrm{a}}$ & \\
\hline
\hline
2 & $48.42 \pm 0.12$ & & $49.21 \pm 0.18$ & & $47.98 \pm 0.17$ & & $48.97 \pm 0.16$ & &
$47.91 \pm 0.17$ & & $56.82 \pm 0.72$ \\
10 & $43.35 \pm 0.07$ & & $52.50 \pm 0.50$ & & $44.56 \pm 0.22$ & & $50.86 \pm 0.45$ & &
$43.19 \pm 0.13$ & & $*$ \\
100 & $38.90 \pm 0.04$ & & $*$ & & $*$ & & $*$ & &
$39.67 \pm 0.14$ & & $*$ \\
\hline
\hline
\end{tabular}
\caption{As in Table \ref{tbl:sim1} with the axial ratio of the elliptical PSF equal to to $1:2$
\label{tbl:sim2}}
\end{center}
\end{table*}
\begin{table*}[t]
\begin{center}
\begin{tabular}{ccccccccccccc}
\hline
\hline
\multicolumn{1}{c}{} &
\multicolumn{1}{c}{{\rm Wiener}} & \multicolumn{1}{c}{} &
\multicolumn{1}{c}{{\rm Ref. Ext.}} & \multicolumn{1}{c}{} &
\multicolumn{1}{c}{{\rm Tikhonov}} & \multicolumn{1}{c}{} &
\multicolumn{1}{c}{{\rm Wind. Ref. Ext.}} & \multicolumn{1}{c}{} &
\multicolumn{1}{c}{{\rm Wind. Tikh.}} & \multicolumn{1}{c}{} &
\multicolumn{1}{c}{{\rm Kron. Approx.}} \\
\cline{2-2} \cline{4-4} \cline{6-6} \cline{8-8} \cline{10-10} \cline{12-12} \\
${\rm S/N}$ & ${\rm rrms\, (\%)}$ & & ${\rm rrms\, (\%)}^{\mathrm{a}} $ & & ${\rm rrms\, (\%)}$ &
& ${\rm rrms\, (\%)}^{\mathrm{a}} $ & &
${\rm rrms\, (\%)}$ & & ${\rm rrms\, (\%)}^{\mathrm{a}}$ & \\
\hline
\hline
2 & $51.04 \pm 0.55$ & & $51.91 \pm 0.56$ & & $50.64 \pm 0.62$ & & $51.85 \pm 0.55$ & &
$50.60 \pm 0.63$ & & $51.17 \pm 0.65$ \\
10 & $45.73 \pm 0.38$ & & $52.81 \pm 1.15$ & & $47.11 \pm 0.74$ & & $51.91 \pm 1.07$ & &
$45.56 \pm 0.41$ & & $*$ \\
100 & $41.12 \pm 0.26$ & & $*$ & & $*$ & & $*$ & &
$42.65 \pm 0.38$ & & $*$ \\
\hline
\hline
\end{tabular}
\caption{As in Table~\ref{tbl:sim1} with the only difference that a new Gaussian random field is generated
for each simulation.
\label{tbl:sim1a}}
\end{center}
\end{table*}
\begin{table*}[t]
\begin{center}
\begin{tabular}{ccccccccccccc}
\hline
\hline
\multicolumn{1}{c}{} &
\multicolumn{1}{c}{{\rm Wiener}} & \multicolumn{1}{c}{} &
\multicolumn{1}{c}{{\rm Ref. Ext.}} & \multicolumn{1}{c}{} &
\multicolumn{1}{c}{{\rm Tikhonov}} & \multicolumn{1}{c}{} &
\multicolumn{1}{c}{{\rm Wind. Ref. Ext.}} & \multicolumn{1}{c}{} &
\multicolumn{1}{c}{{\rm Wind. Tikh.}} & \multicolumn{1}{c}{} &
\multicolumn{1}{c}{{\rm Kron. Approx.}} \\
\cline{2-2} \cline{4-4} \cline{6-6} \cline{8-8} \cline{10-10} \cline{12-12} \\
${\rm S/N}$ & ${\rm rrms\, (\%)}$ & & ${\rm rrms\, (\%)}^{\mathrm{a}} $ & & ${\rm rrms\, (\%)}$ &
& ${\rm rrms\, (\%)}^{\mathrm{a}} $ & &
${\rm rrms\, (\%)}$ & & ${\rm rrms\, (\%)}^{\mathrm{a}}$ & \\
\hline
\hline
2 & $48.63 \pm 0.50$ & & $49.99 \pm 0.51$ & & $48.32 \pm 0.57$ & & $49.75 \pm 0.50$ & &
$48.28 \pm 0.57$ & & $57.20 \pm 1.11$ \\
10 & $43.23 \pm 0.33$ & & $52.61 \pm 1.32$ & & $44.12 \pm 0.54$ & & $51.52 \pm 1.13$ & &
$43.10 \pm 0.36$ & & $*$ \\
100 & $38.62 \pm 0.22$ & & $*$ & & $*$ & & $*$ & &
$39.33 \pm 0.27$ & & $*$ \\
\hline
\hline
\end{tabular}
\caption{As in Table~\ref{tbl:sim2} with the only difference that a new Gaussian random field is generated
for each simulation.
\label{tbl:sim2a}}
\end{center}
\end{table*}

\section{Numerical experiments} \label{sec:numerical}
\subsection{A preliminary experiment}

\begin{figure}
        \resizebox{\hsize}{!}{\includegraphics{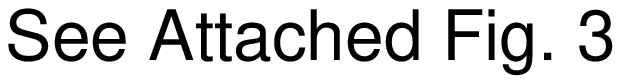}}
        \caption{Grayscale image of the central $290 \times 290$ pixels of a simulated sky map at the {\it
          PLANCK}-LFI
          frequencies with ${\rm S/N}=2$ and axial ratio of the elliptical PSF equal to $1:1.3$ (see text).
        The original map was of
        $354 \times 354$ pixels, i.e., a border of $32$ pixels has been removed from each side
          of the image, corresponding to about four times the dispersion of the PSF along the major axis.
        $N_w = 32$ has been used for the dewindowed modified Hanning methods as well as {\it Periodic}
          BC and discrete Laplacian for $\Lb$. For the approximated Kronecker method a {\it reflexive} BC
        has been adopted.}
        \label{fig:debia_67_2}
\end{figure}
\begin{figure}
        \resizebox{\hsize}{!}{\includegraphics{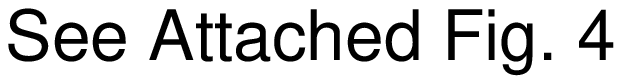}}
        \caption{As in Fig.~\ref{fig:debia_67_2} but with ${\rm S/N}=10$ and axial ratio $ = 1:1.3$.}
        \label{fig:debia_67_10}
\end{figure}
\begin{figure}
        \resizebox{\hsize}{!}{\includegraphics{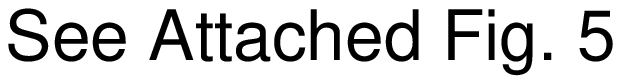}}
        \caption{As in Fig.~\ref{fig:debia_67_2} but with ${\rm S/N} = 2$ and axial ratio $= 1:2$.}
        \label{fig:debi_67_2}
\end{figure}
\begin{figure}
        \resizebox{\hsize}{!}{\includegraphics{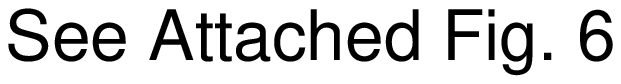}}
        \caption{As in Fig.~\ref{fig:debia_67_10} but with ${\rm S/N} = 10$ and axial ratio $= 1:2$.}
        \label{fig:debi_67_10}
\end{figure}

To support our arguments and study the performance of the approximations and methods described in the previous
sections, we present the result of numerical simulations
based on a Gaussian random process with statistical characteristics similar to those expected for the CMB emission
at the frequencies typical of {\it PLANCK}-LFI (see VNT). The PSF is Gaussian with elliptical symmetry,
the {\rm FWHM} along the major axis is $\approx 8$ pixels (about two times the worst spatial resolution
expected for {\it PLANCK}), and with axes forming an angle of $45^{\circ}$ with the edges of the map.
We consider the ${\rm S/N}$ ratios $2$, $10$, $100$; and two values of the axial ratio: $1:1.3$ and $1:2$.
The axial ratios and the width of
the PSF used in the experiment are by far less favorable than those
expected for {\it PLANCK} and represent a sort of {\it worst possible scenario}. The reason for using
${\rm S/N}$ ratios much more larger than the value expected for {\it PLANCK} ($\approx 2$) is that
some characteristics of the method are hidden by high noise contamination.
Six deblurring methods are tested: classic Wiener, Tikhonov (with
periodic BC and discrete Laplacian for $\Lb$) applied both
to $\gbh$ and $\gbt$ (windowed Tikhonov), reflexive extension method (with periodic BC and discrete Laplacian for $\Lb$)
applied both to a windowed and to an unwindowed extended image, and the Kronecker approximation
with reflexive BC.
Since the random field is Gaussian and stationary and the noise is assumed
white, classical Wiener filtering is expected to provide the smallest mean
square error among linear filters. However, this filter requires knowledge of the spectrum of the unknown
signal which is not available in practice. We use Wiener deblurring, based on the real spectrum of the signal, as a
sort of benchmark to assess the performance of Tikhonov methods. For this reason, it has also been
implemented in a way that avoids edge effects.

The simulations have been conducted under two different scenarios. We first fix the sky and generate different
realizations of the noise process. Then, to account for the variability of the random field, we simulate different
realizations of the random field and the noise process.

\begin{figure}
        \resizebox{\hsize}{!}{\includegraphics{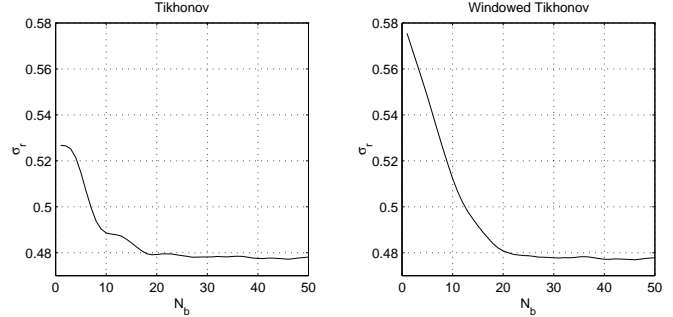}}
        \caption{Standard deviation $\sigma_r$ of the residual between the deblurred and the true maps vs.
          the width $N_b$ of the removed border for the map in Fig.~\ref{fig:debi_67_2}}.
        \label{fig:deb_67_2}
\end{figure}

\begin{figure}
        \resizebox{\hsize}{!}{\includegraphics{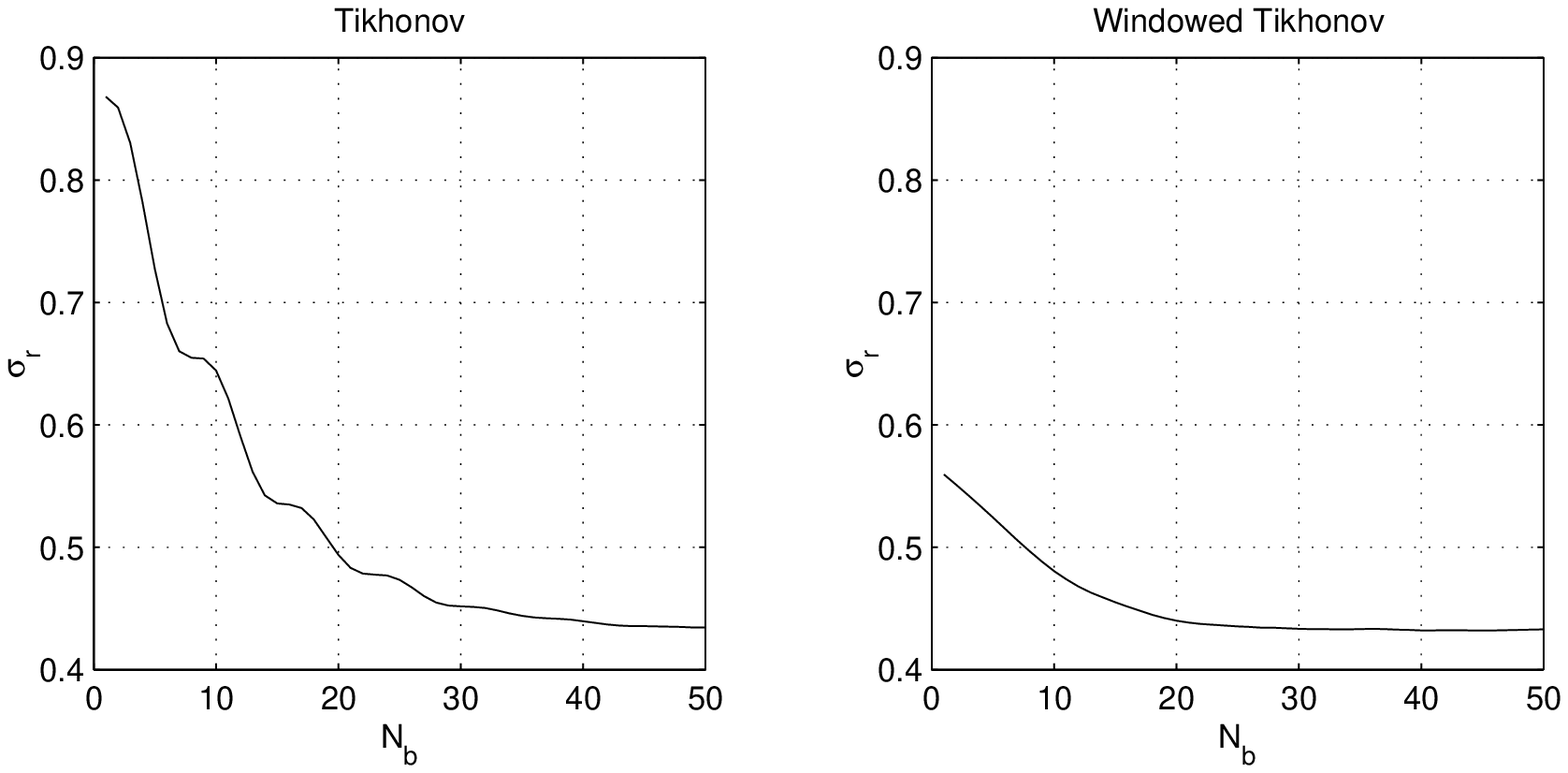}}
        \caption{Standard deviation $\sigma_r$ of the residual between the deblurred and the true maps vs.
          the width $N_b$ of the removed border for the map in Fig.~\ref{fig:debi_67_10}}.
       \label{fig:deb_67_10}
\end{figure}

\begin{figure}
        \resizebox{\hsize}{!}{\includegraphics{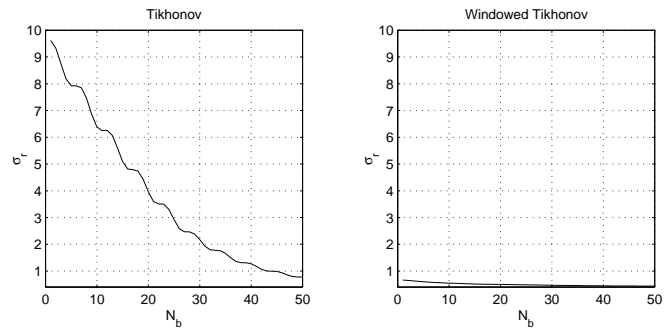}}
        \caption{As in Figs.~\ref{fig:deb_67_2} and \ref{fig:deb_67_10} but with ${\rm S/N}=100$}.
        \label{fig:deb_67_100}
\end{figure}

Tabs.\ref{tbl:sim1}-\ref{tbl:sim2a} and Figs.\ref{fig:debia_67_2}-\ref{fig:debi_67_10} confirm that the Kronecker
approximation method does not perform well with a very asymmetric PSF. Furthermore, it is evident that the windowed
Tikhonov method performs the best, close to the Wiener filter, even in the case of very high ${\rm S/N}$
that is troublesome for the other methods. Figs.~\ref{fig:deb_67_2}-\ref{fig:deb_67_100} show
the typical behavior of the standard deviation of the residuals of the deblurred and the true maps
as larger borders are removed from the frames. The effect of the Gibbs phenomenon is evident
in the figures, especially those obtained by directly deblurring the unwindowed $\gbh$ and for
high values of ${\rm S/N}$. These figures indicate that for moderate ${\rm S/N}$ ratios
a border of thickness $3$-$4$ times the dispersion of the PSF has to be removed after the deblurring
to reduce edge effects. By increasing $N_w$, this method still performs better even for very high
${\rm S/N}$ ratios. In typical CMB applications the ${\rm S/N}$ is low, thus here we do not consider
the question of finding an ``optimal'' value of $N_w$. In fact, our simulations indicate that
a value of $N_w$ equal to $3$-$4$ times the dispersion of the PSF along the major axis is a reasonable choice.
This value corresponds approximately to the thickness of the border in which the blurred
image is influenced by data outside of the field of view.

Finally, as shown in Tables \ref{tbl:sim1}-\ref{tbl:sim2a}, the methods provide similar results for low
${\rm S/N}$ ratios, especially when a sufficiently large number of edge pixels is removed from the image.
This is not surprising as edge effects remain close to the edges and high noise levels hide the
effects of the smallest eigenvalues of $\Hb$ that cause
the ill-posedness of the deblurring operation.


Tables \ref{tbl:sim3} and \ref{tbl:sim4} show the results of $100$ simulations
similar to those presented previously
but under the conditions expected for {\it PLANCK}-LFI. As expected, given the low ${\rm S/N}$ ratio,
the performance of the various methods is similar.

\begin{table*}[t]
\begin{center}
\begin{tabular}{ccccccccccc}
\hline
\hline
\multicolumn{1}{c}{}
& \multicolumn{2}{c}{{\rm Wiener}} & \multicolumn{1}{c}{}
& \multicolumn{3}{c}{{\rm Windowed Tikhonov}} & \multicolumn{1}{c}{}
& \multicolumn{3}{c}{{\rm Kronecker Approx.}} \\
\cline{3-3} \cline{5-7} \cline{9-11}\\
${\rm FWHM\,(arc.min.)}$ & &${\rm rrms\, (\%)}$ & & ${\rm rrms\, (\%)}$ & & $\lambda$
& & ${\rm rrms\, (\%)}^{\mathrm{a}}$ & & $\lambda$ \\
\hline\hline
$10$ & & $30.36 \pm 0.06$ & & $30.35 \pm 0.06$ & & $0.69 \pm 0.02$ & & $30.45 \pm 0.05$ & & $0.69 \pm 0.01$\\
$14$ & & $32.64 \pm 0.07$ & & $36.68 \pm 0.09$ & & $0.79 \pm 0.04$ & & $37.96 \pm 0.08$ & & $0.79 \pm 0.04$\\
$23$ & & $36.74 \pm 0.08$ & & $36.68 \pm 0.09$ & & $0.79 \pm 0.04$ & & $36.96 \pm 0.08$ & & $0.79 \pm 0.02$\\
$33$ & & $40.54 \pm 0.10$ & & $40.47 \pm 0.11$ & & $0.77 \pm 0.04$ & & $40.70 \pm 0.11$ & & $0.79 \pm 0.02$\\
\hline\hline
\end{tabular}
\caption{Summary of the results concerning the deblurring of a sky map contaminated with $100$ realizations
of a Gaussian random process whose statistical properties are similar to those expected of the CMB sky
observed with four channels of {\it PLANCK}-LFI for beams with elliptical symmetry.
The central $320 \times 320$ pixels of the images are considered.
The original maps consisted of $350 \times 350$ pixels (corresponding to a sky area of about
$20^\circ \times 20^\circ$), i.e., a border of $15$ pixels has been removed from
each side of the images. The techniques used are Wiener filtering, windowed Tikhonov
(with periodic boundary conditions and discrete Laplacian for $\Lb$) and the Kronecker approximation
(with reflexive boundary conditions and discrete Laplacian for $\Lb$). Here ${\rm S/N=2}$,
${\rm FWHM}$ is the full width at half maximum along the major axis of the PSF,
the axial ratio for the PSF is $1:1.3$ and $N_w=15$. The relative {\it root mean square} ({\rm rrms}) is
defined as the ratio of the residual root mean square ({\rm rms}) to the rms of the true signal.
When applicable, the mean values and dispersions of the GCV estimates of $\lambda$ are also shown.
\label{tbl:sim3}}
\end{center}
\vspace*{\newsp}
\begin{list}{}{}
\item[$^{\mathrm{a}}$] Calculated on the entire images of $350 \times 350$ pixels.
\end{list}
\end{table*}

\begin{table*}[t]
\begin{center}
\begin{tabular}{ccccccccccc}
\hline
\hline
\multicolumn{1}{c}{}
& \multicolumn{2}{c}{{\rm Wiener}} & \multicolumn{1}{c}{}
& \multicolumn{3}{c}{{\rm Windowed Tikhonov}} & \multicolumn{1}{c}{}
& \multicolumn{3}{c}{{\rm Kronecker Approx.}} \\
\cline{3-3} \cline{5-7} \cline{9-11}\\
${\rm FWHM\,(arc.min.)}$ & &${\rm rrms\, (\%)}$ & & ${\rm rrms\, (\%)}$ & & $\lambda$
& & ${\rm rrms\, (\%)}^{\mathrm{a}}$ & & $\lambda$ \\
\hline\hline
$10$ & & $30.68 \pm 0.26$ & & $30.71 \pm 0.29$ & & $0.73 \pm 0.03$ & & $30.73 \pm 0.28$ & & $0.73 \pm 0.03$\\
$14$ & & $32.68 \pm 0.24$ & & $32.68 \pm 0.28$ & & $0.75 \pm 0.04$ & & $32.73 \pm 0.27$ & & $0.76 \pm 0.04$\\
$23$ & & $36.74 \pm 0.24$ & & $36.67 \pm 0.28$ & & $0.78 \pm 0.04$ & & $36.78 \pm 0.26$ & & $0.79 \pm 0.04$\\
$33$ & & $40.65 \pm 0.27$ & & $40.52 \pm 0.30$ & & $0.77 \pm 0.05$ & & $40.73 \pm 0.29$ & & $0.81 \pm 0.05$\\
\hline\hline
\end{tabular}
\caption{As in Table~\ref{tbl:sim3} with the only difference that a new Gaussian random
field is generated for each simulation.
\label{tbl:sim4}}
\end{center}
\end{table*}

\subsection{Numerical simulations with realistic CMB maps}

The next step is to test the performance of the deblurring procedure
with realistic CMB maps from the point of view of the angular power spectrum.
As usual, it is defined as the set of coefficients $C_{\ell}$ of
the two-point correlation function expanded in Legendre polynomials;
$\ell$ marks the power at the angular scale given approximately
by $\theta\simeq 180/\ell$. We compare the quality of the reconstruction
with the corresponding one in VNT, which assumed a circular Gaussian beam.

We consider the same templates as in VNT: the region
is a squared patch ($350 \times 350$ pixels) with side
of about $20^{\circ}$, centered
at $l=90^{\circ}$, $b=45^{\circ}$ (Galactic coordinates). The latitude
is high enough that CMB emission dominates over foregrounds, assumed
to be represented by synchrotron \citep{has82} and dust \citep{sch98}
emission. We neglect contributions of point
sources. The CMB model, in agreement with current experimental results
\citep{deb02, hal02, lee01},
corresponds to a flat Friedmann-Robertson-Walker (FRW) metric
with a cosmological constant ($70\%$ of the critical density),
Hubble parameter today $H_{0}=100h$ km/sec/Mpc with $h=0.7$
baryons at $5\%$ and Cold Dark Matter (25\% CDM), with
a scale-invariant Gaussian initial spectrum of adiabatic
density perturbations.

The {\it PLANCK}-LFI instrument works at frequencies
$30$, $44$, $70$, and $100~{\rm GHz}$. We assume nominal
noise and angular resolution corresponding to the four frequencies
$30$, $44$, $70$, and $100~{\rm GHz}$ at which the {\it PLANCK}-LFI instrument works.
The simulated maps are blurred through Gaussian PSF's with elliptical symmetry. In particular,
the following FWHM's along the major axis have been used: $\approx 33^{\prime}$ at $30 ~{\rm GHz}$,
$\approx 23^{\prime}$ at $44 ~{\rm GHz}$, $\approx 14^{\prime}$ at $70 ~{\rm GHz}$,
$\approx 10^{\prime}$ at $100 ~{\rm GHz}$, with axial ratio set to $1:1.3$, and axes forming
an angle of $45^{\circ}$ with the edges of the map. Simulated
white noise, with {\rm rms} level as expected for the considered channels,
has been added to the maps.
Since we choose to work with a pixel size of about $3.5$ arcminutes, the noise
rms are $.042$, $.049$, $.042$ and $.043$ mK in antenna temperature at $30$, $44$, $70$,
$100~{\rm GHz}$, respectively.

Again, as in VNT, we can see two important characteristics
of the deblurring process: first it reconstructs the
correct shape and amplitude of the part of the spectrum which
is mildly affected by the PSF; second, it reconstructs part of
the power where the signal is degraded substantially by the
PSF and noise. In fact, we see that the performance of the deblurring method is almost the
same as that reported in VNT for symmetric PSF.
Indeed, in the $30$, $44$, $70$ and $100 ~{\rm GHz}$ cases the spectrum is reconstructed
up to $\ell\simeq 400$, $500$, $700$ and $800$ respectively.

We conclude that, for what concerns the power spectrum analysis, the proposed technique seems to work well also
with asymmetric shapes of the instrumental beam.

Note also that for a fixed $\lambda$ the Tikhonov estimate (\ref{eq:tikhonov}) is a
linear function of the data and therefore the covariance matrix of the deblurred field can be obtained by 
propagating the covariance matrix of the original field through the linear operators. Even if the GCV 
estimate of $\lambda$ makes the estimate nonlinear, for large samples the linear approximation that 
assumes $\lambda$ is fixed is reasonable. In any case, it is necessary to stress that
the most feasible way to account for the changes in the statistics of the CMB through
any estimation process (to include the effects of, for example, edge effects, possible unremoved
point sources, asymmetry and/or non-stationary PSF etc.), is probably through
Monte Carlo simulations where one applies the same algorithm to the data and to simulated maps to 
compare and fit the best cosmological model. The low computational cost of our algorithm is certainly 
interesting with this respect.

\section{Discussion and conclusions} \label{sec:final}
We have considered Tikhonov regularization for deblurring CMB maps in real space. As shown in VNT,
this approach permits the development of algorithms that are more flexible and robust than those based on
frequency-space methods. The methods developed in VNT, however, apply to the case of symmetric
PSFs for which efficient methods can be implemented with reflexive BC. In the present paper
we have considered the more general case of asymmetric PSFs.

We have presented a method based
on a Kronecker separable approximation of the PSF that can be used with mildly asymmetric PSFs.
For more asymmetric cases we presented a periodic BC approach that can be efficiently implemented
with image windows and fast Fourier transforms. Windowing is necessary to reduce edge effect in
the selection of the regularization parameter. Of course, one can easily derive a GCV function that
only takes into account pixels away from the edges but the computational cost is higher.

We have applied our methodology to
simulated skies at typical CMB frequencies.
We considered test signals with known statistics,
as well as realistic simulations of the CMB sky
contaminated by noise whose rms is that expected for the low frequency
instrument aboard the {\it PLANCK} satellite.
This case is particularly interesting for
application of a deblurring procedure, as the
instrument observes the sky at 30, 44, 70 and 100 GHz
with very different PSFs of resolution $33$, $22$, $14$, $10$
arcminutes. We found that the proposed methodology performs as well or better than the Wiener
benchmark that relies on the true spectrum and that avoids edge effects.

\begin{figure}
        \resizebox{\hsize}{!}{\includegraphics{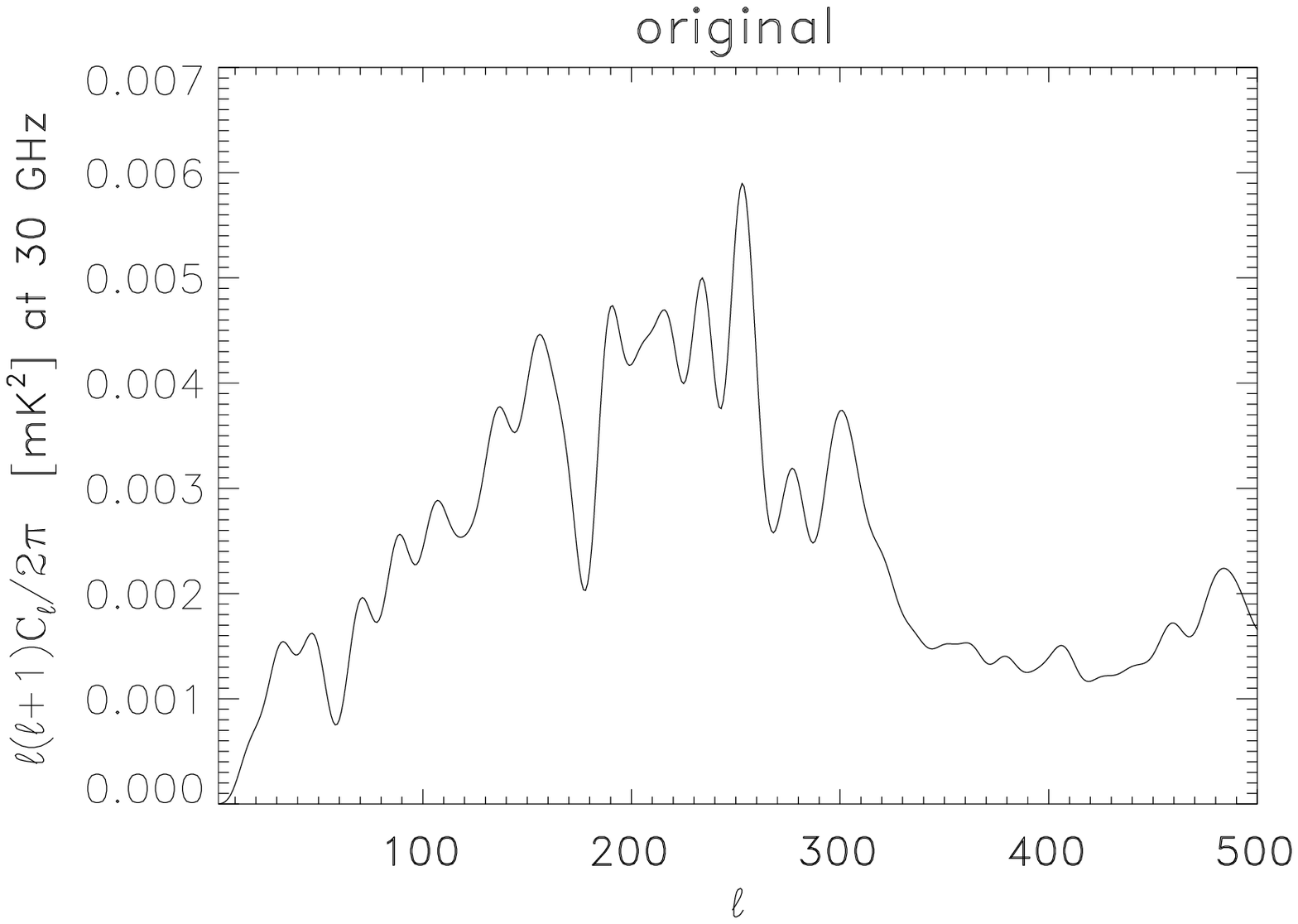}\includegraphics{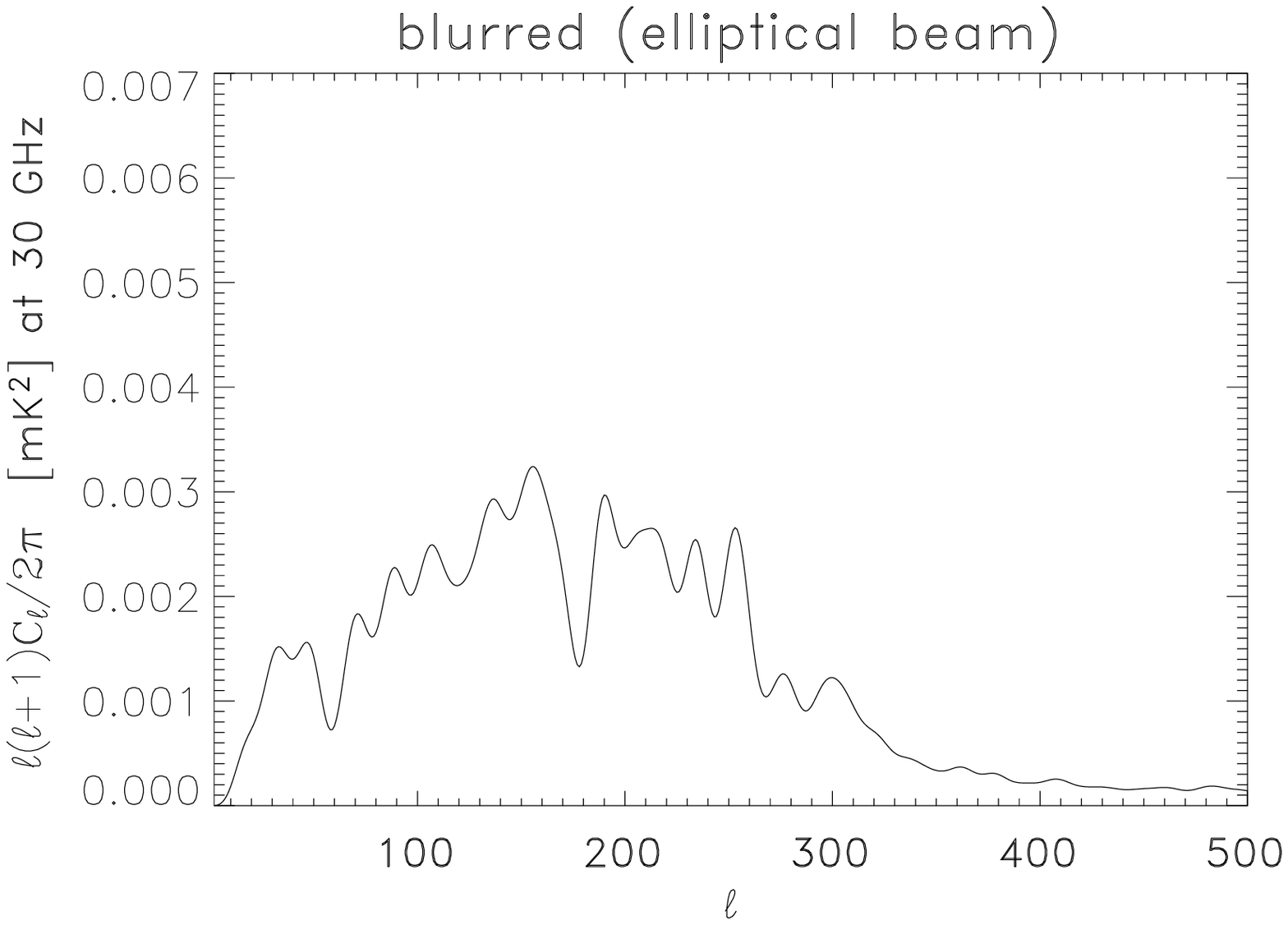}}
        \resizebox{\hsize}{!}{\includegraphics{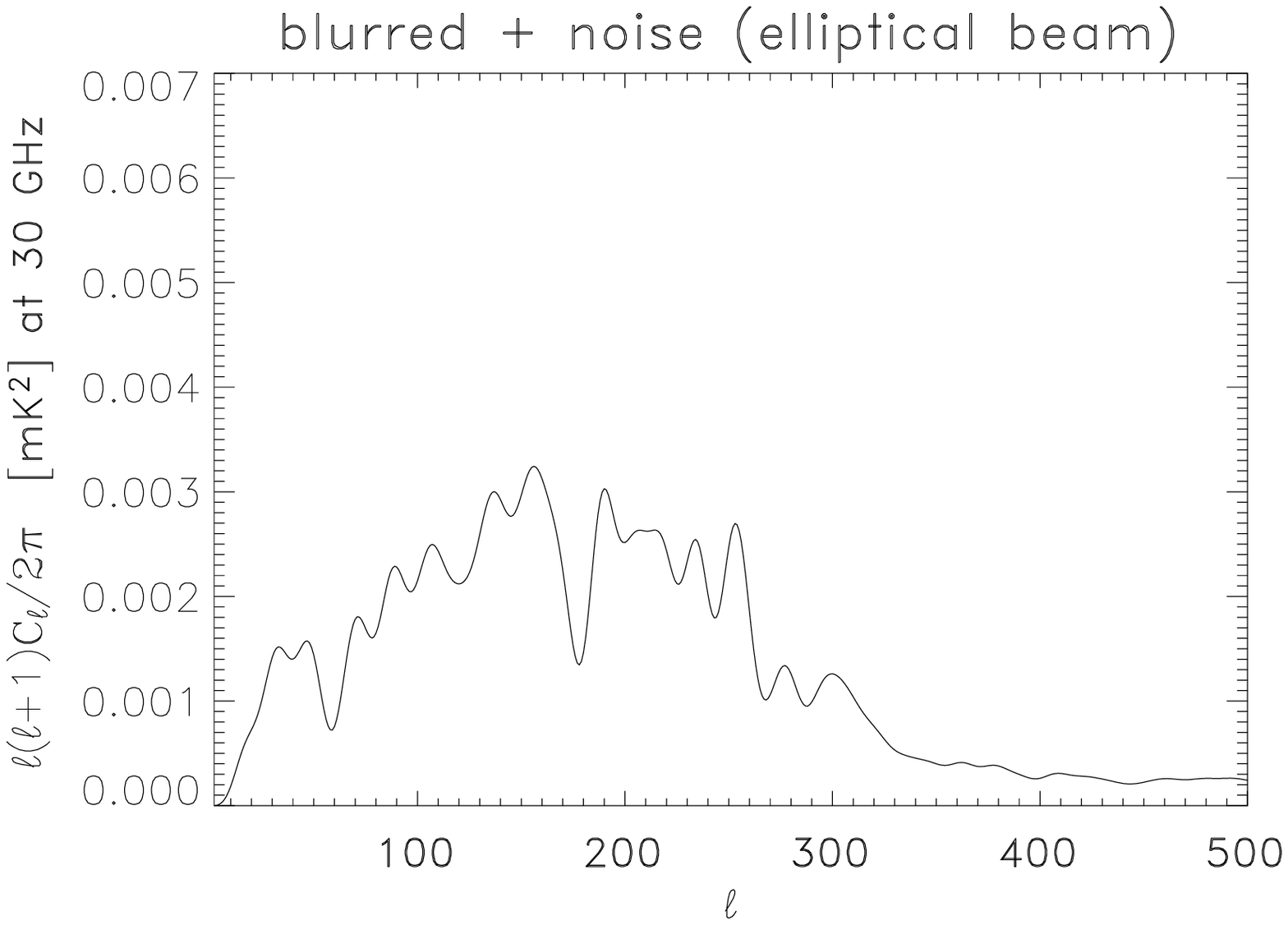}\includegraphics{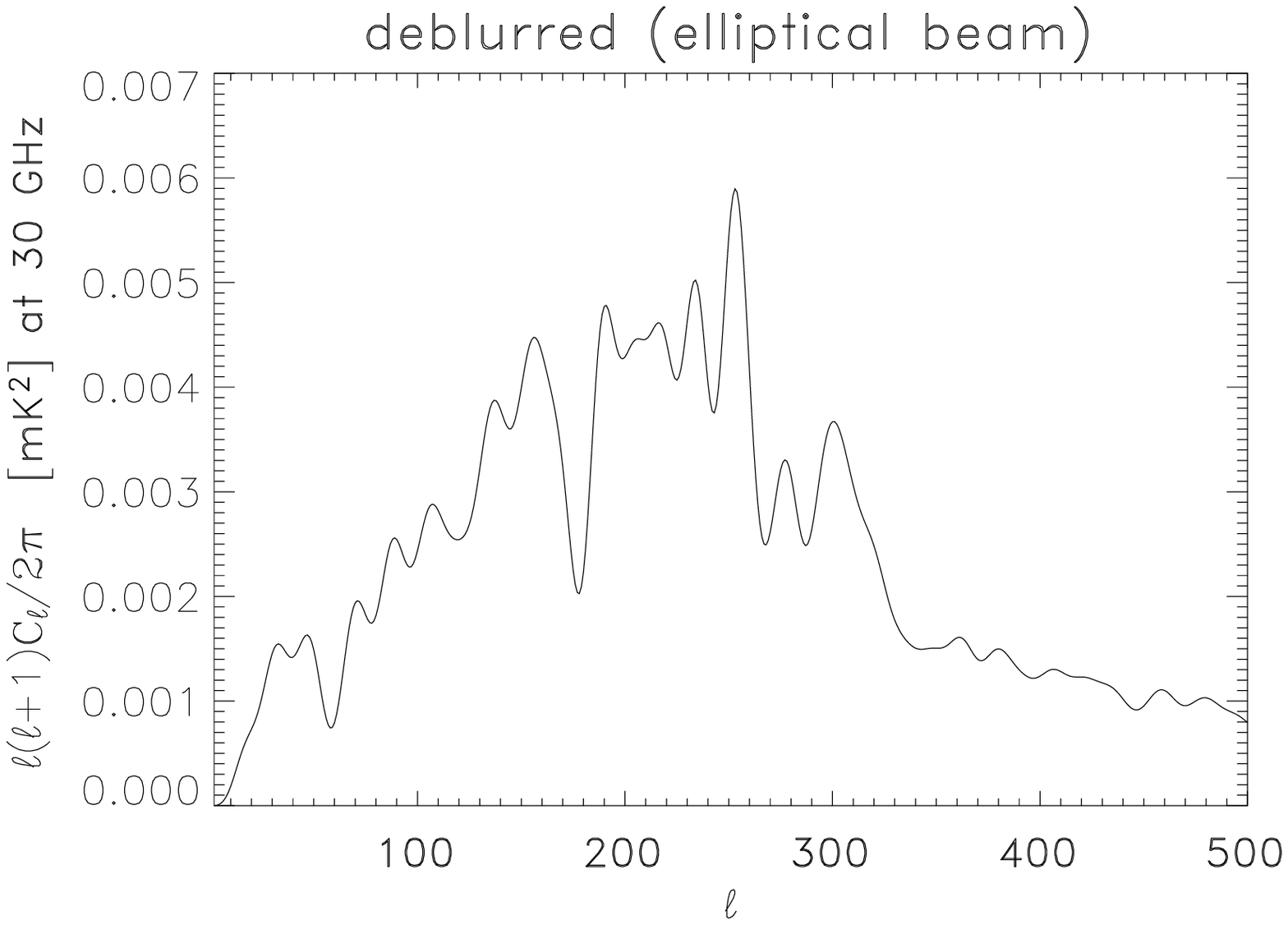}}
        \caption{Angular power spectrum at $30 ~{\rm GHz}$ in different steps of the analysis.}
        \label{fig:ps30}
\end{figure}
\begin{figure}
        \resizebox{\hsize}{!}{\includegraphics{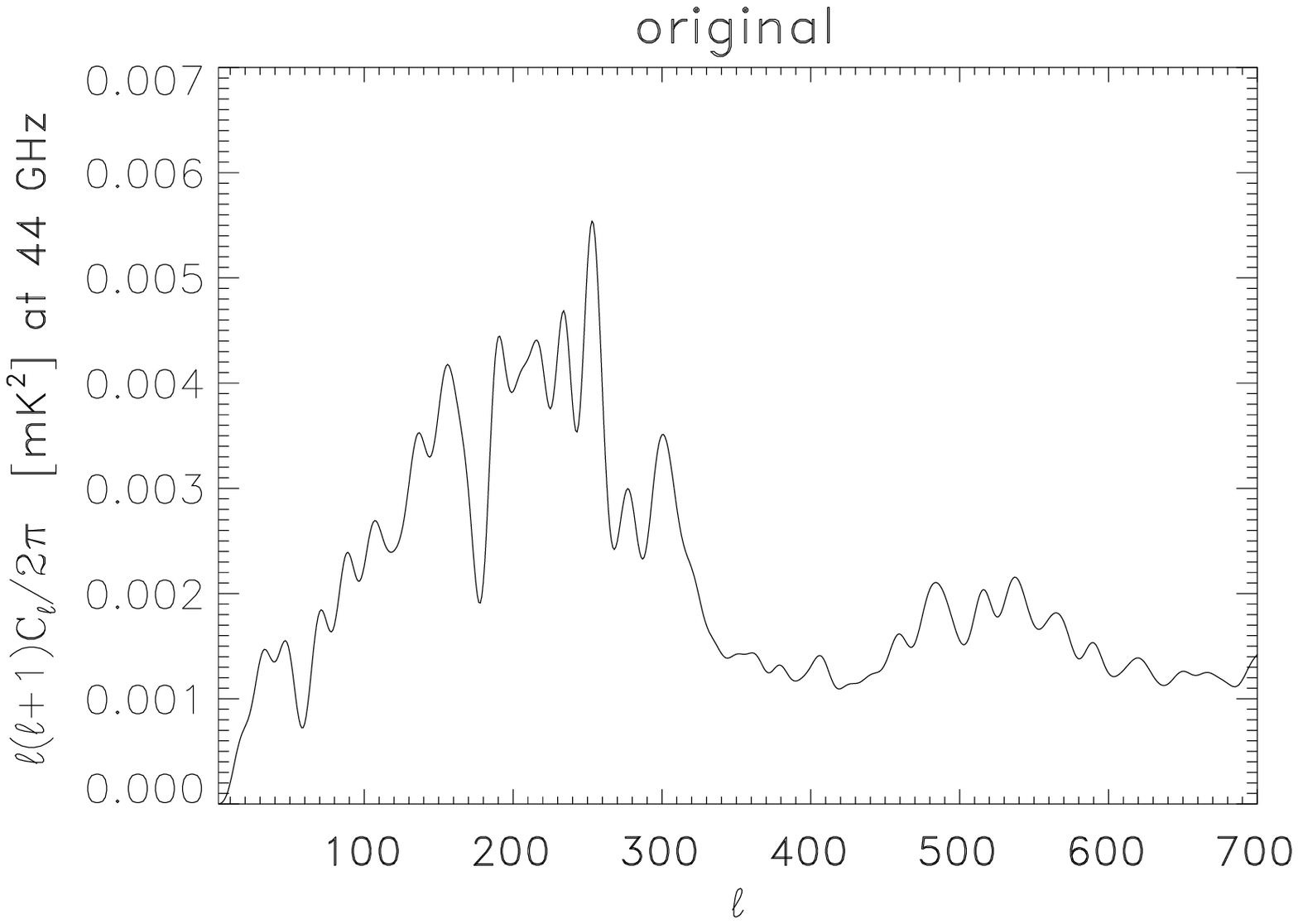}\includegraphics{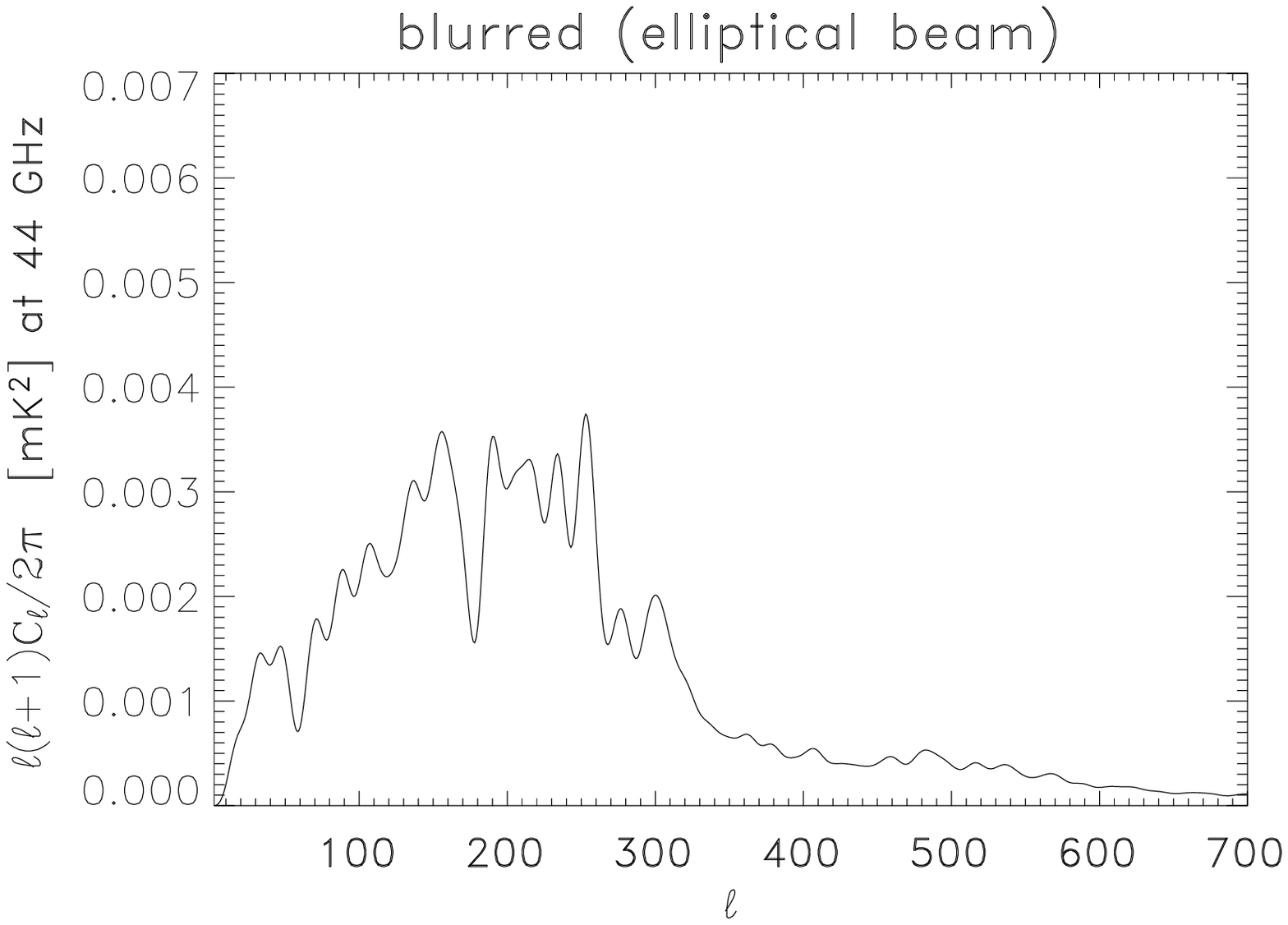}}
        \resizebox{\hsize}{!}{\includegraphics{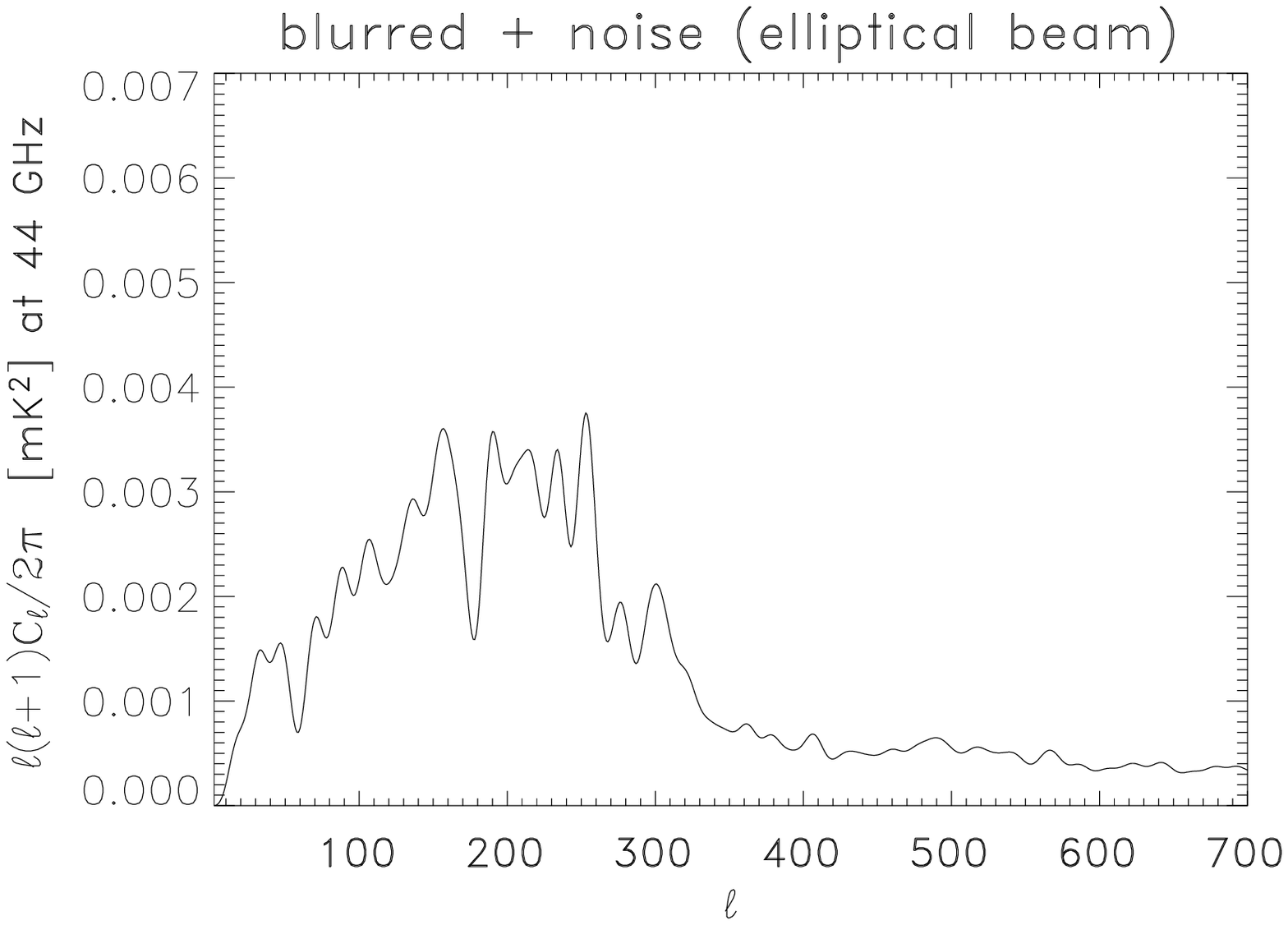}\includegraphics{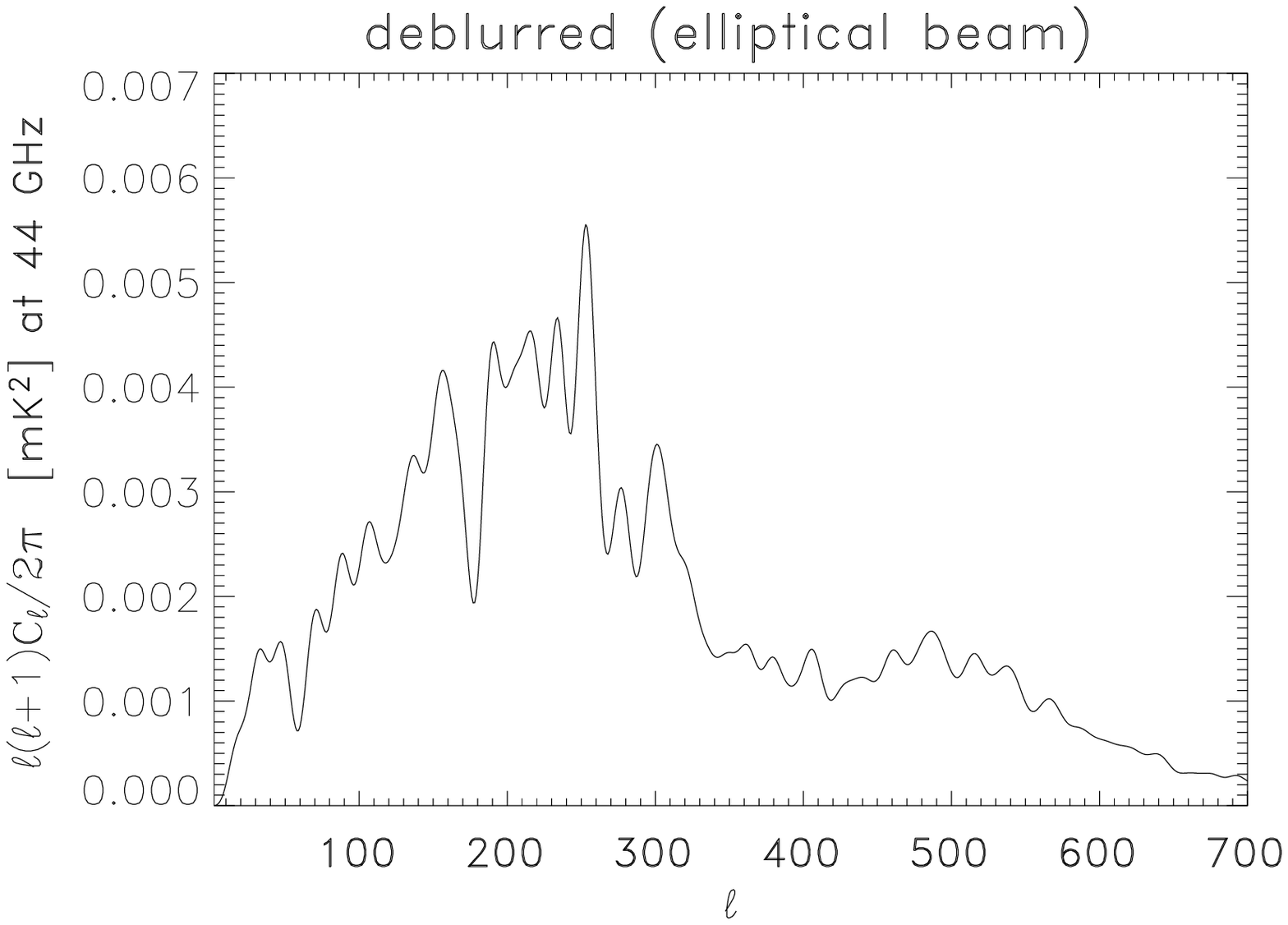}}
        \caption{Angular power spectrum at $44 ~{\rm GHz}$ in different steps of the analysis.}
        \label{fig:ps44}
\end{figure}
\begin{figure}
        \resizebox{\hsize}{!}{\includegraphics{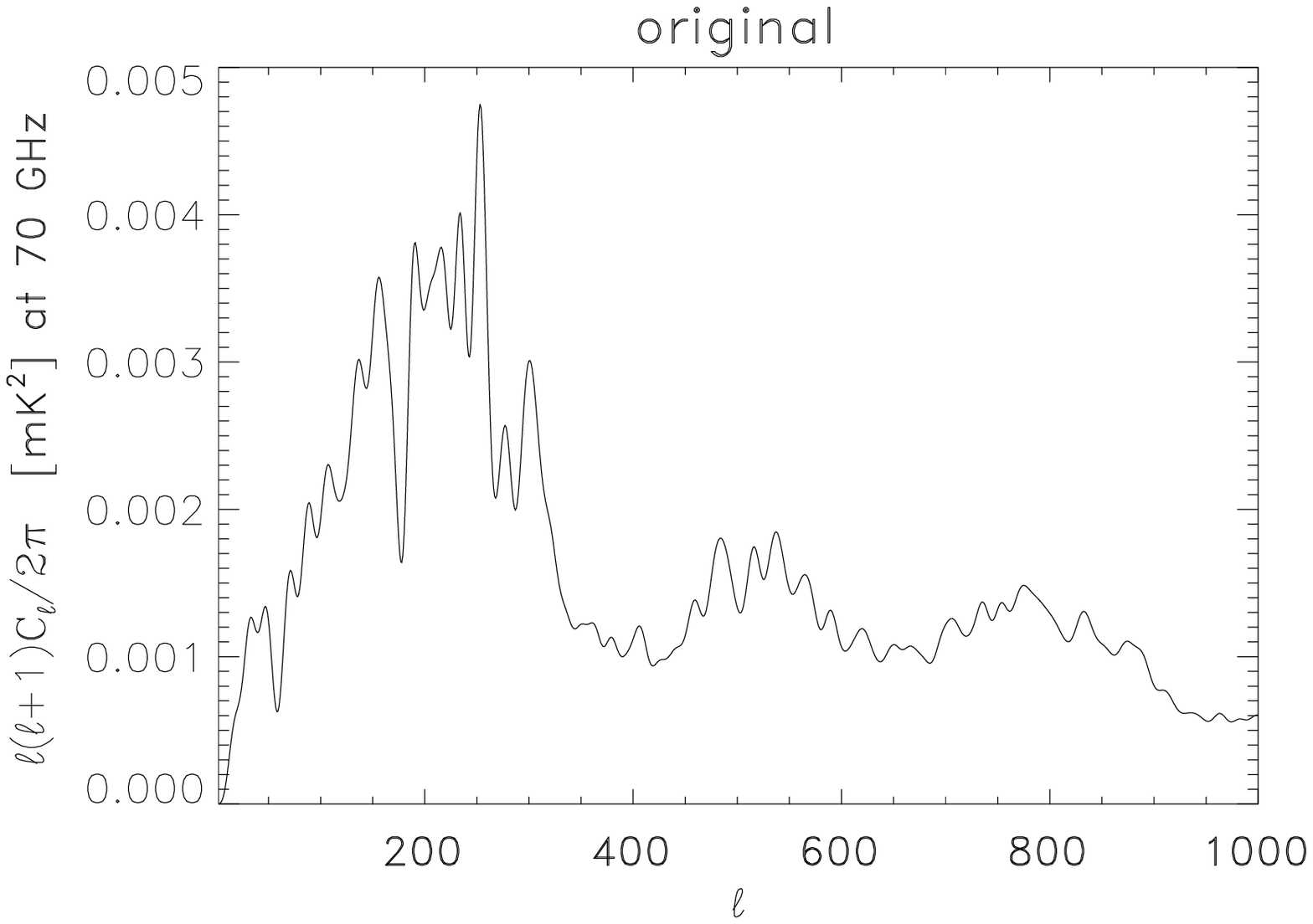}\includegraphics{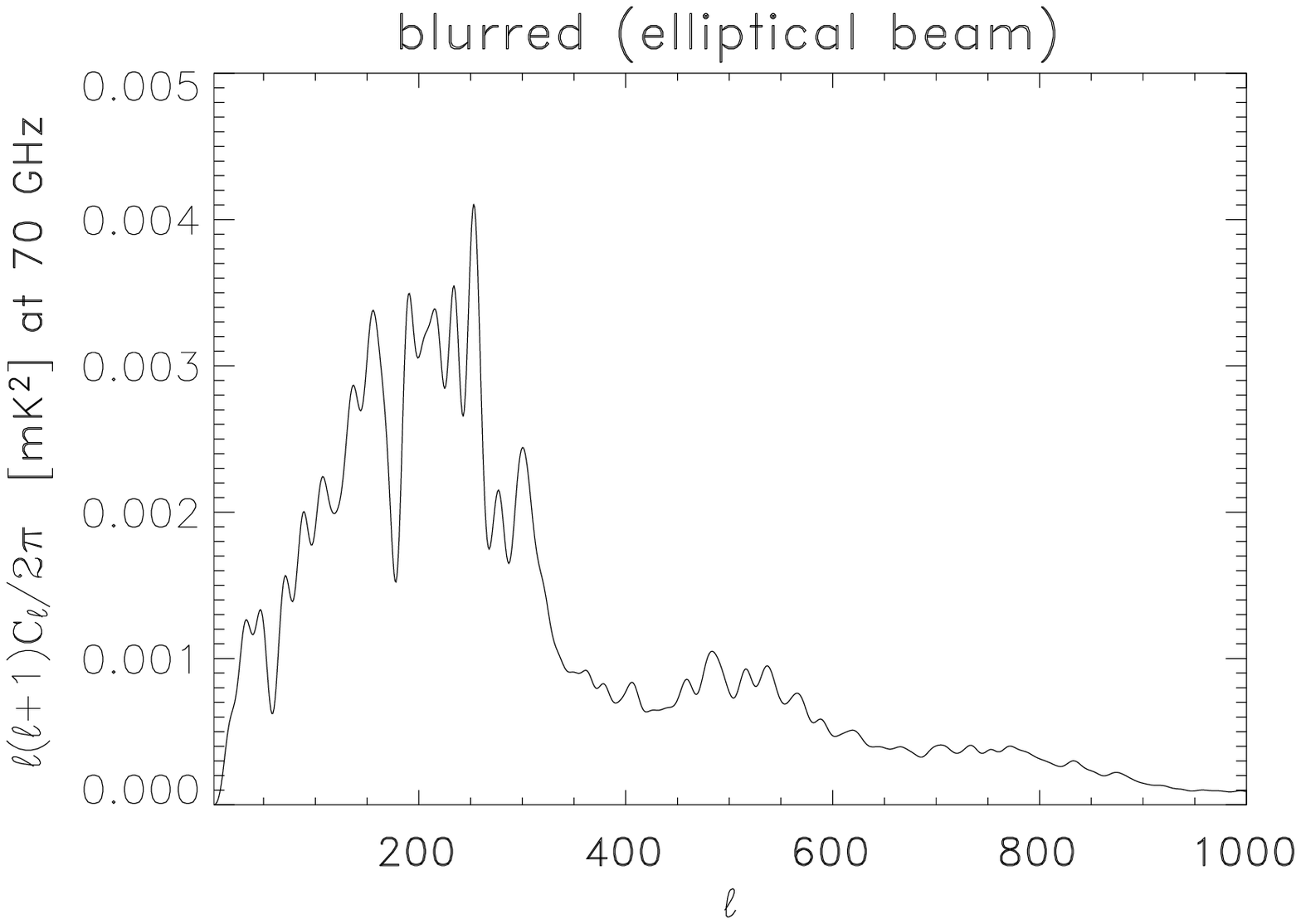}}
        \resizebox{\hsize}{!}{\includegraphics{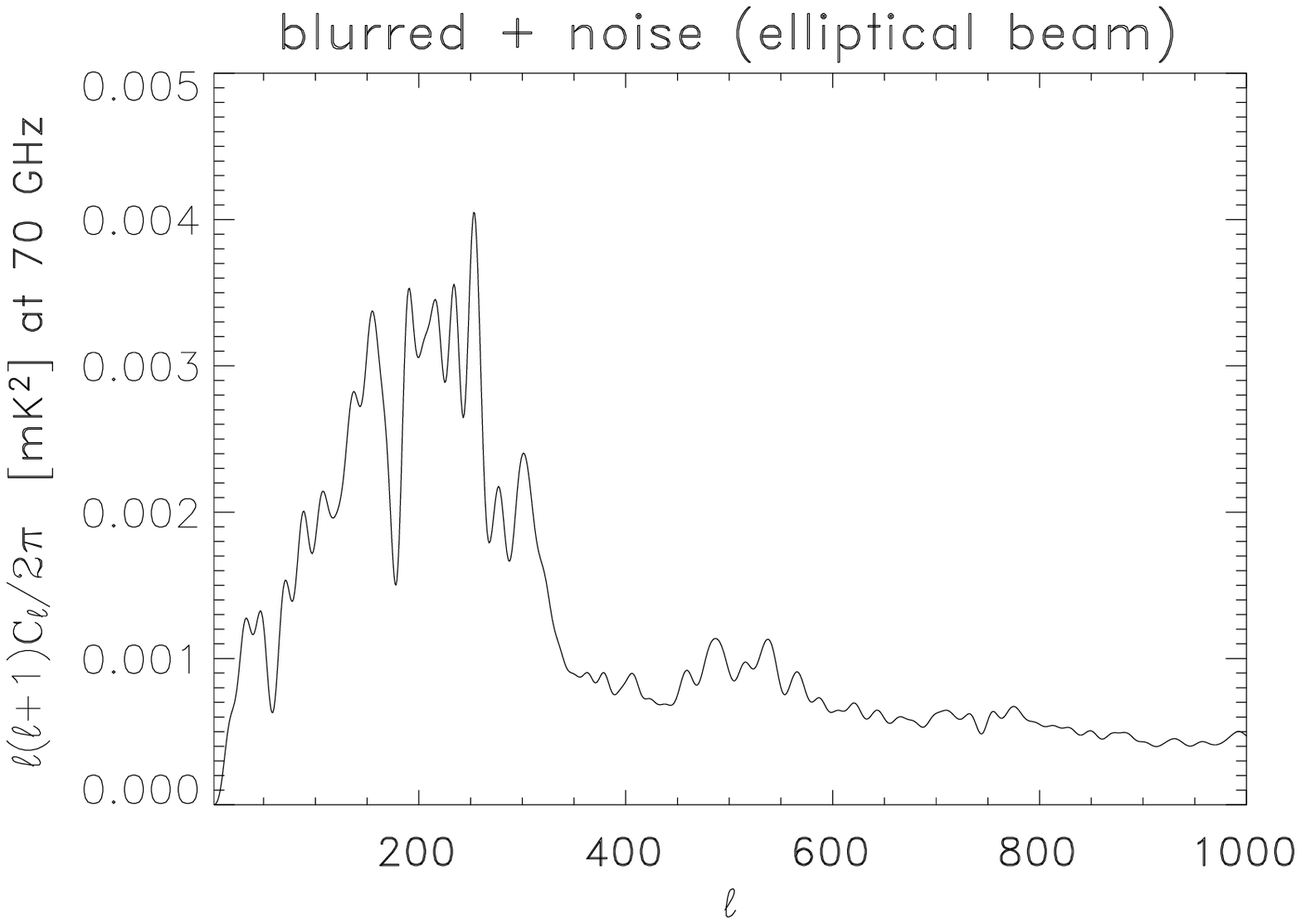}\includegraphics{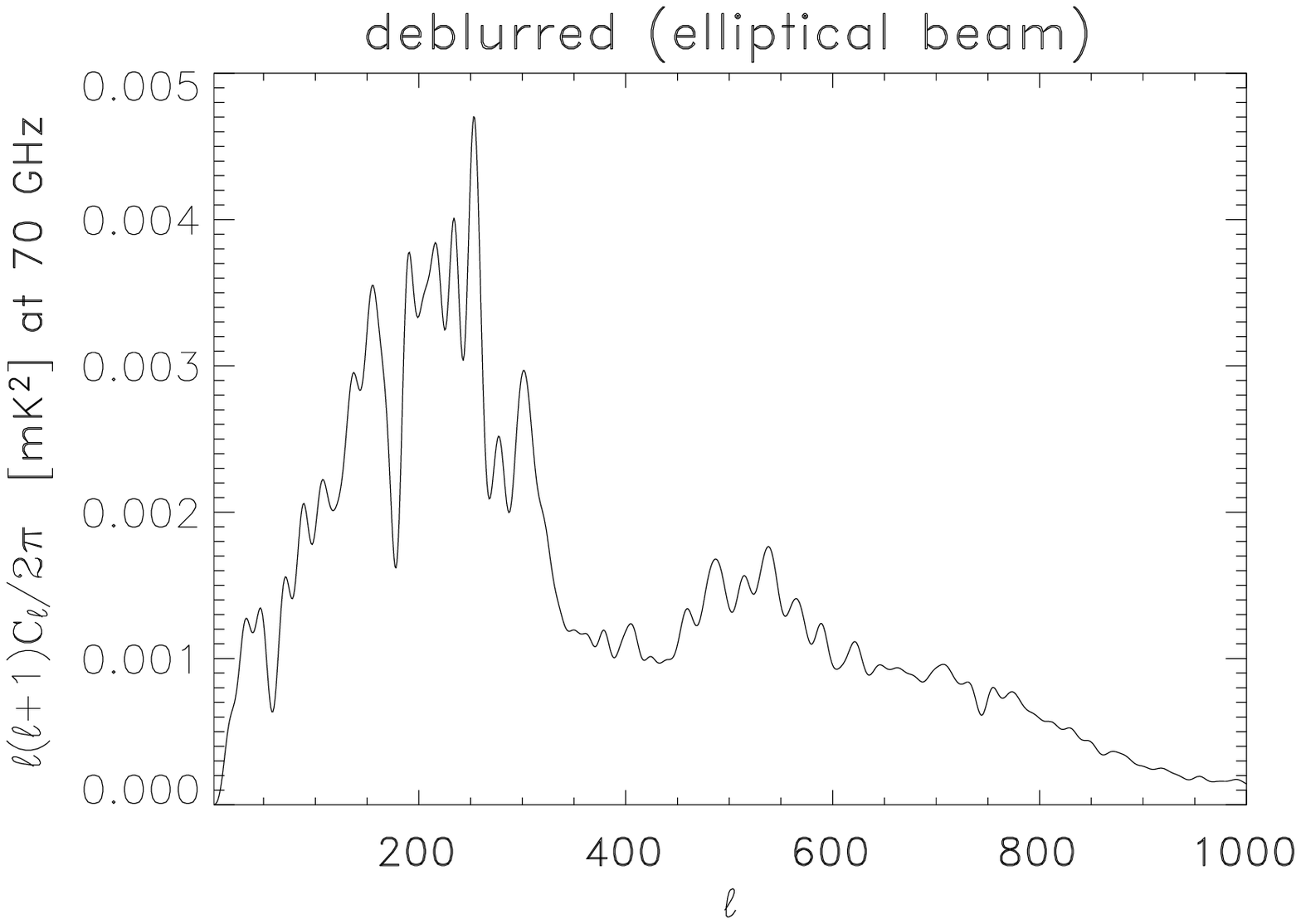}}
        \caption{Angular power spectrum at $70 ~{\rm GHz}$ in different steps of the analysis.}
        \label{fig:ps70}
\end{figure}
\begin{figure}
        \resizebox{\hsize}{!}{\includegraphics{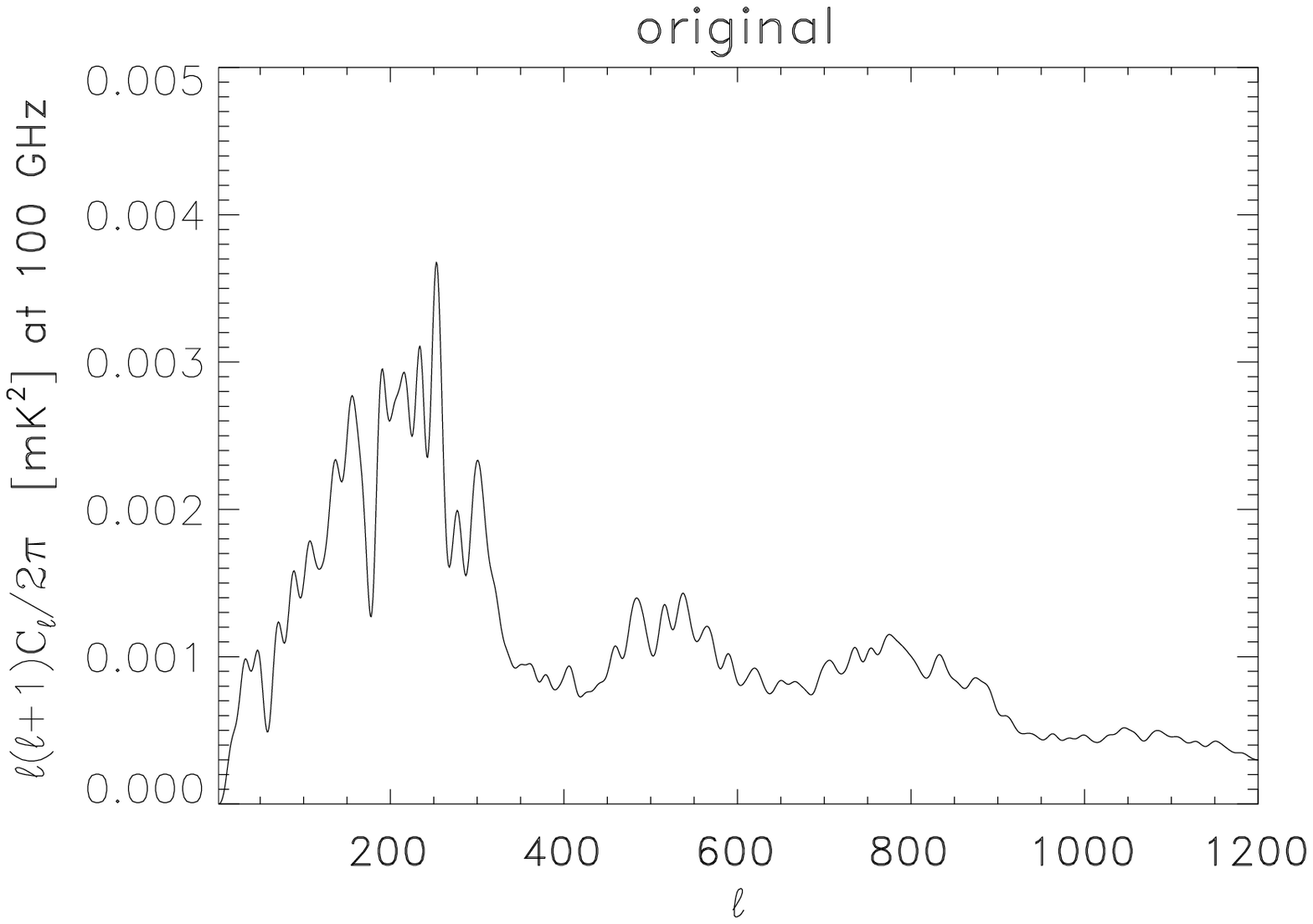}\includegraphics{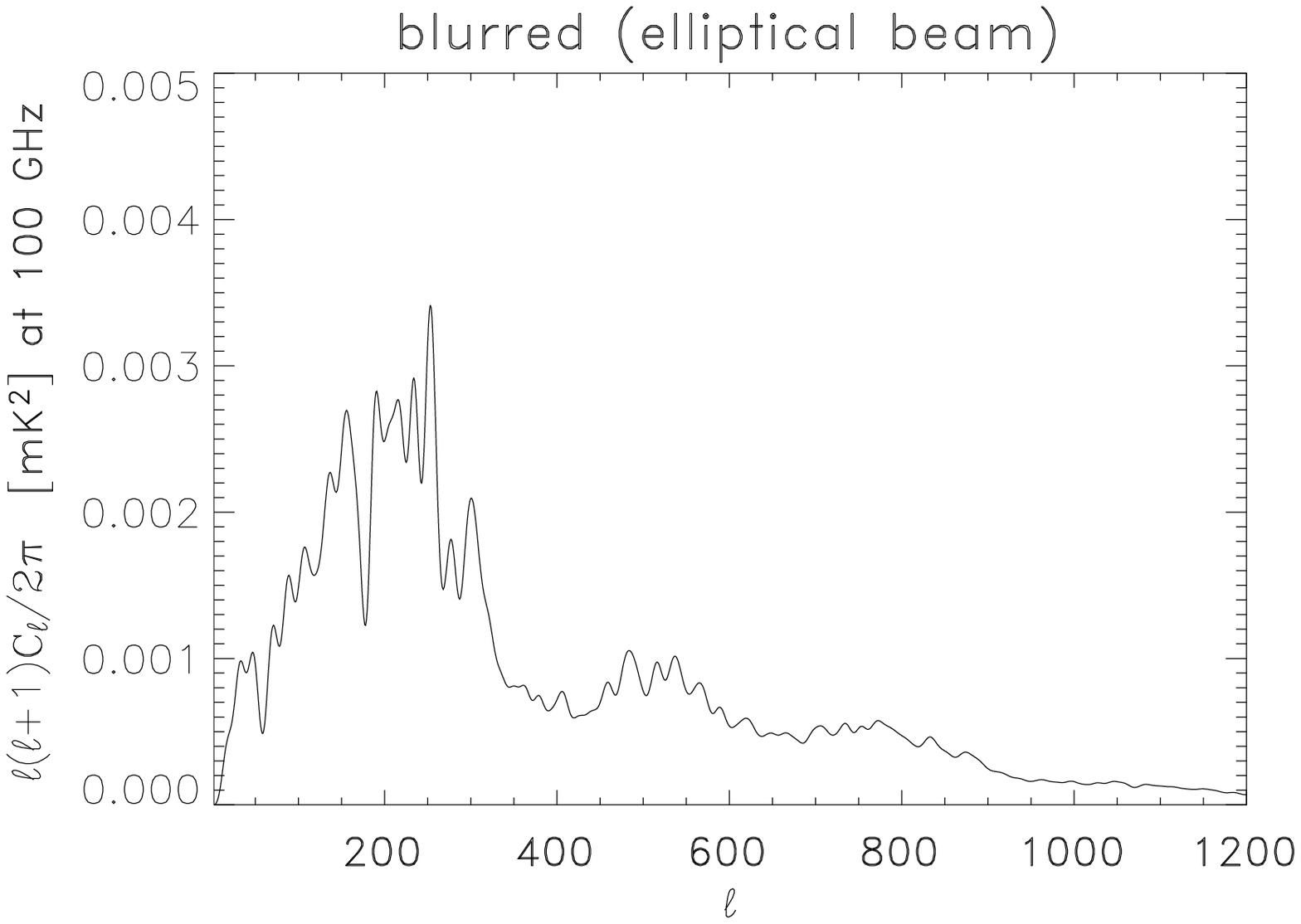}}
        \resizebox{\hsize}{!}{\includegraphics{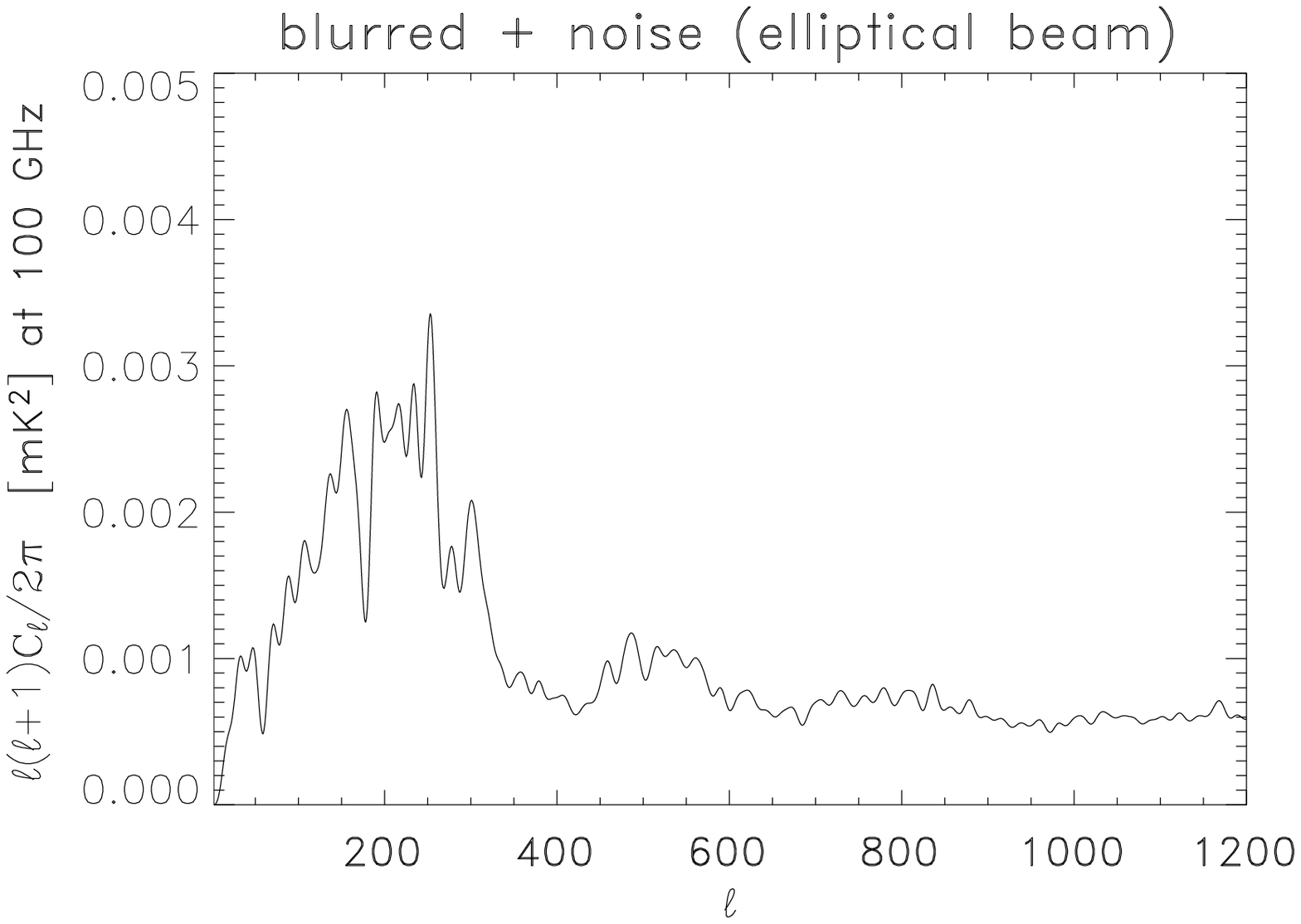}\includegraphics{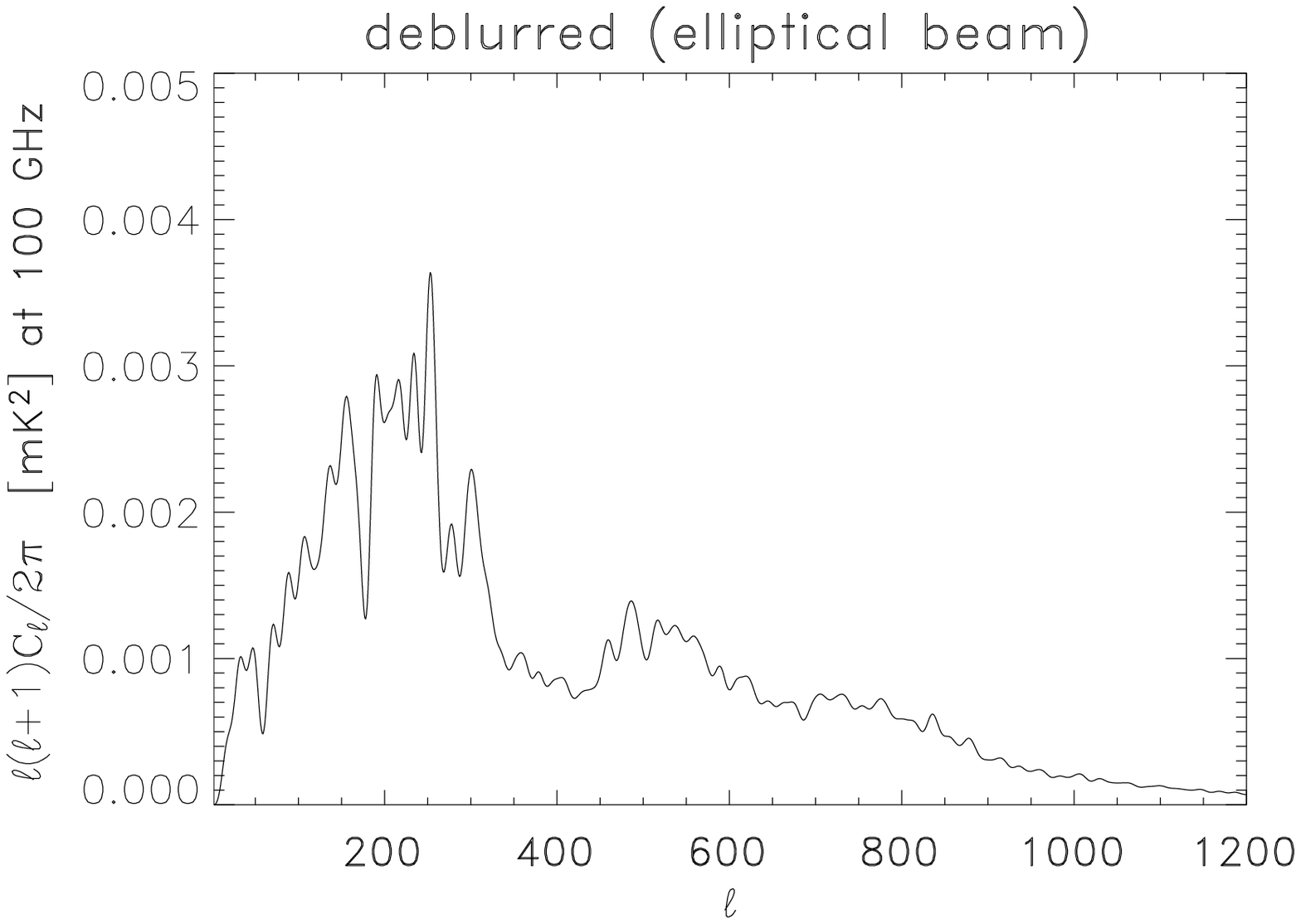}}
        \caption{Angular power spectrum at $100 ~{\rm GHz}$ in different steps of the analysis.}
        \label{fig:ps100}
\end{figure}


\begin{thebibliography}{}
\bibitem[Baccigalupi et al. (2000)]{bac00} Baccigalupi, C., Bedini, L., Burigana, C. et al. 2000, MNRAS, 318, 769
\bibitem[de Bernardis et al. (2002)]{deb02} de Bernardis, P., Ape, P.A.R., \& Bock, J.J. 2002, ApJ, 564, 559
\bibitem[Golub \& Van Loan (1996)]{GoVa96} Golub, G.H., \& Van Loan, C.F. 1996,
Matrix Computations, 3rd ed. (Johns Hopkins, Baltimore)
\bibitem[Halverson et al. (2002)]{hal02} Halverson, N.W., Leitch, E.M, \& Pryke, C. 2002, ApJ, 568, 38
\bibitem[Haslam et al. (1982)]{has82} Haslam, C.G.T., Stoffel, H., Salter, C.J., \& Wilson, W.E. 1982, A\&AS, 47, 1
\bibitem[Kamm \& Nagy (2000)]{KaNa00} Kamm, J., \& Nagy, J.G. 2000,
SIAM J. Matrix Anal. and Appl., 22, 155
\bibitem[Lee et al. (2001)]{lee01} Lee, B.C., Tucker, D.L. et al. 2001, ApJL, 561, 183
\bibitem[Maino et al. (2002)]{mai02} Maino, D., Farusi, A., Baccigalupi, C. et al. 2002, MNRAS, 334, 53
\bibitem[Nagy et al. (2002)]{NaNgPe02} Nagy, J.G., Ng, M.K., \& Perrone, L. 2002,
SIAM J. Matrix Anal. and Appl., {\it submitted}
\bibitem[Schlegel et al. (1998)]{sch98} Schlegel, D.J., Finkbeiner, D.P., \& Davies, M. 1998, ApJ, 500, 525
\bibitem[Van~Loan \& Pitsianis (1993)]{VLoPi93} Van~Loan, C.F. \& Pitsianis, N.P. 1993
in Moonen, M.S. \& Golub, G.H. ed., Linear Algebra for Large Scale and Real Time Applications,
Kluwer Publications, 293
\bibitem[Tegmark (1999)]{teg99} Tegmark, M. 1999, ApJ, 519, 513
\bibitem[Vio et al. (2003)]{vio03} Vio, R., Nagy, J.G., Tenorio, L. et al. 2003, A\&A, 389, 404 (VNT)
\end{thebibliography}
\end{document}